\PassOptionsToPackage{unicode}{hyperref}
\PassOptionsToPackage{hyphens}{url}
\PassOptionsToPackage{dvipsnames,svgnames,x11names}{xcolor}
\documentclass[
  12pt]{article}

\usepackage{amsmath,amssymb}
\usepackage{iftex}
\ifPDFTeX
  \usepackage[T1]{fontenc}
  \usepackage[utf8]{inputenc}
  \usepackage{textcomp} 
\else 
  \usepackage{unicode-math}
  \defaultfontfeatures{Scale=MatchLowercase}
  \defaultfontfeatures[\rmfamily]{Ligatures=TeX,Scale=1}
\fi
\usepackage{lmodern}
\ifPDFTeX\else  
\fi
\IfFileExists{upquote.sty}{\usepackage{upquote}}{}
\IfFileExists{microtype.sty}{
  \usepackage[]{microtype}
  \UseMicrotypeSet[protrusion]{basicmath} 
}{}
\makeatletter
\@ifundefined{KOMAClassName}{
  \IfFileExists{parskip.sty}{%
    \usepackage{parskip}
  }{
    \setlength{\parindent}{0pt}
    \setlength{\parskip}{6pt plus 2pt minus 1pt}}
}{
  \KOMAoptions{parskip=half}}
\makeatother
\usepackage[table]{xcolor}
\makeatletter
\ifx\paragraph\undefined\else
  \let\oldparagraph\paragraph
  \renewcommand{\paragraph}{
    \@ifstar
      \xxxParagraphStar
      \xxxParagraphNoStar
  }
  \newcommand{\xxxParagraphStar}[1]{\oldparagraph*{#1}\mbox{}}
  \newcommand{\xxxParagraphNoStar}[1]{\oldparagraph{#1}\mbox{}}
\fi
\ifx\subparagraph\undefined\else
  \let\oldsubparagraph\subparagraph
  \renewcommand{\subparagraph}{
    \@ifstar
      \xxxSubParagraphStar
      \xxxSubParagraphNoStar
  }
  \newcommand{\xxxSubParagraphStar}[1]{\oldsubparagraph*{#1}\mbox{}}
  \newcommand{\xxxSubParagraphNoStar}[1]{\oldsubparagraph{#1}\mbox{}}
\fi
\makeatother

\usepackage{longtable,booktabs,array}
\usepackage{calc} 
\usepackage{etoolbox}
\makeatletter
\patchcmd\longtable{\par}{\if@noskipsec\mbox{}\fi\par}{}{}
\makeatother
\IfFileExists{footnotehyper.sty}{\usepackage{footnotehyper}}{\usepackage{footnote}}
\makesavenoteenv{longtable}
\usepackage{graphicx}
\makeatletter
\def\maxwidth{\ifdim\Gin@nat@width>\linewidth\linewidth\else\Gin@nat@width\fi}
\def\maxheight{\ifdim\Gin@nat@height>\textheight\textheight\else\Gin@nat@height\fi}
\makeatother
\setkeys{Gin}{width=\maxwidth,height=\maxheight,keepaspectratio}
\makeatletter
\def\fps@figure{htbp}
\makeatother

\makeatletter
\@ifpackageloaded{caption}{}{\usepackage{caption}}
\AtBeginDocument{%
\ifdefined\contentsname
  \renewcommand*\contentsname{Table of contents}
\else
  \newcommand\contentsname{Table of contents}
\fi
\ifdefined\listfigurename
  \renewcommand*\listfigurename{List of Figures}
\else
  \newcommand\listfigurename{List of Figures}
\fi
\ifdefined\listtablename
  \renewcommand*\listtablename{List of Tables}
\else
  \newcommand\listtablename{List of Tables}
\fi
\ifdefined\figurename
  \renewcommand*\figurename{Figure}
\else
  \newcommand\figurename{Figure}
\fi
\ifdefined\tablename
  \renewcommand*\tablename{Table}
\else
  \newcommand\tablename{Table}
\fi
}
\@ifpackageloaded{float}{}{\usepackage{float}}
\floatstyle{ruled}
\@ifundefined{c@chapter}{\newfloat{codelisting}{h}{lop}}{\newfloat{codelisting}{h}{lop}[chapter]}
\floatname{codelisting}{Listing}

\makeatother
\makeatletter
\@ifpackageloaded{caption}{}{\usepackage{caption}}
\@ifpackageloaded{subcaption}{}{\usepackage{subcaption}}
\makeatother

\ifLuaTeX
  \usepackage{selnolig}  
\fi
\usepackage[]{natbib}
\usepackage{bibunits}
\defaultbibliographystyle{agsm}
\defaultbibliography{bibliography}
\usepackage{bookmark}

\IfFileExists{xurl.sty}{\usepackage{xurl}}{} 
\hypersetup{
  pdftitle={Statistical Validation of Computer Models: Global and Subdomain Hypothesis Testing},
  pdfkeywords={Function comparison, Uncertainty quantification, Computer experiments},
  colorlinks=true,
  linkcolor={blue},
  filecolor={Maroon},
  citecolor={Blue},
  urlcolor={Blue},
  pdfcreator={LaTeX via pandoc}}

\usepackage{amsfonts, amsthm, latexsym}
\usepackage{bm}

\theoremstyle{plain}
\newtheorem{theorem}{Theorem}[section]
\newtheorem{proposition}{Proposition}[section]
\newtheorem{lemma}{Lemma}[section]
\theoremstyle{remark}
\newtheorem{remark}{Remark}[section]
\usepackage{mathtools}
\usepackage{algorithmic}
\usepackage{algorithm}

\usepackage{cleveref}
\usepackage{enumitem}
\usepackage{multirow}
\usepackage{mathrsfs}
\usepackage{epstopdf}

\newcommand{\anon}{1}
\if1\anon\hypersetup{pdfauthor={Chaoan Li; Xianyang Zhang; Rui Tuo}}\fi 

\newcommand{\markrevision}{0}
\if1\markrevision\definecolor{revcolor}{named}{red}\else\definecolor{revcolor}{named}{black}\fi
\newcommand{\new}[1]{{\color{revcolor}#1}}
\newenvironment{newmaterial}{\color{revcolor}}{}

\newcommand{\showlinenumbers}{0}
\if1\showlinenumbers
\usepackage[pagewise]{lineno}
\newcommand*\patchAmsMathEnvironmentForLineno[1]{%
  \expandafter\let\csname old#1\expandafter\endcsname\csname #1\endcsname
  \expandafter\let\csname oldend#1\expandafter\endcsname\csname end#1\endcsname
  \renewenvironment{#1}{\linenomath\csname old#1\endcsname}{\csname oldend#1\endcsname\endlinenomath}}
\newcommand*\patchBothAmsMathEnvironmentsForLineno[1]{%
  \patchAmsMathEnvironmentForLineno{#1}\patchAmsMathEnvironmentForLineno{#1*}}
\AtBeginDocument{%
  \patchBothAmsMathEnvironmentsForLineno{equation}%
  \patchBothAmsMathEnvironmentsForLineno{align}%
  \patchBothAmsMathEnvironmentsForLineno{eqnarray}%
  \patchBothAmsMathEnvironmentsForLineno{gather}%
  \linenumbers}
\else
\makeatletter\providecommand\@LN[2]{}\makeatother 
\fi

\begin{document}

\def\spacingset#1{\renewcommand{\baselinestretch}%
{#1}\small\normalsize} \spacingset{1}


\if1\anon
{
  \title{\bf Statistical Validation of Computer Models: Global and Subdomain Hypothesis Testing}
  \author{Chaoan Li\thanks{
    The authors acknowledge the generous support from NSF grants DMS-2312173 and CNS-2328395.}\hspace{.2cm}\\
    Department of Industrial and Systems Engineering, Texas A\&M University\\
    and \\
    Xianyang Zhang \\
    Department of Statistics, Texas A\&M University\\
    and \\
    Rui Tuo \\
    Department of Industrial and Systems Engineering, Texas A\&M University}
  \maketitle
} \fi

\if0\anon
{
  \bigskip
  \bigskip
  \bigskip
  \begin{center}
    {\LARGE\bf Statistical Validation of Computer Models: Global and Subdomain Hypothesis Testing}
\end{center}
  \medskip
} \fi

\bigskip
\begin{abstract}
Computer simulations play an important role in scientific discovery and engineering innovation. Reliable computer models enable virtual experimentation that reduces the need for costly and time-consuming physical testing. However, the credibility of such models hinges on rigorous statistical validation against real-world data. This paper develops a formal frequentist framework for both global and subdomain validation of computer models. We propose the Fourier Maximum Modulus Test (FMMT), which leverages kernel ridge regression (KRR) to estimate the discrepancy between the computer model and the physical process, followed by a frequency-domain test based on weighted generalized Fourier coefficients. The theoretical analysis establishes the asymptotic normality of these coefficients, \new{also when calibration parameters are estimated,} allowing for closed-form p-values. Simulation studies and a shear-layer experiment demonstrate that FMMT achieves competitive power, accurate Type I error control, and, through the subdomain tests, strong sensitivity to localized discrepancies.
\end{abstract}

\noindent%
\vfill

\newpage
\spacingset{1.8} 

\section{Introduction}
\label{sec:intro}


Computer simulations play a significant role in today's scientific discoveries and engineering advancements. Accurate computer models can serve as digital replicas of physical realities, enabling virtual experimentation without the need for physical infrastructure, equipment, and operations. The validation of computer simulation models is a critical step in ensuring their reliability and accuracy before they are employed in practical applications. By rigorously comparing the outcomes of simulations with experimental or real-world data, validation helps confirm that the models accurately represent the physical phenomena they are intended to simulate. This process not only builds confidence in the models' predictive capabilities but also identifies potential limitations and areas for improvement. Effective validation is essential for establishing the credibility of simulations in scientific research, engineering design, and policy-making, where decisions often rely heavily on complex computational models.


A computer model provides a mapping between the input and output variables. The goal of validation is to determine whether this mapping agrees with the \textit{physical process}, defined as the expected value of the real-world outcome as a function of the input. 
In other words, rather than merely comparing model predictions at selected input points, validation should ideally assess whether the functional mapping between inputs and outputs is consistent across the entire input space.

Validation methods span both frequentist and Bayesian frameworks. Frequentist approaches include Hotelling's $T^2$ test \citep{balci1982validation}, engineering frameworks for computational fluid dynamics \citep{oberkampf2002verification,oberkampf2006measures}, and likelihood- or norm-based metrics \citep{hills2002statistical,hills2006model}; these quantify aggregate error but lack formal function-level hypothesis tests. Bayesian approaches model outputs as stochastic processes \citep{sacks1989} and incorporate calibration via explicit discrepancy terms \citep{kennedy2001bayesian,bayarri2007framework,wang2009bayesian}, but do not target formal testing with controlled error rates.

This work fills this gap by formulating the null hypothesis as a functional statement: the computer model agrees with the physical process over the entire input domain. Additionally, we propose subdomain validation on user-specified subregions, because a model may perform adequately globally while failing locally. \new{A rejection shows that the data reveal a discrepancy, and the subdomain tests show where to refine or calibrate the model; whether the discrepancy matters depends on its estimated size and on the accuracy that the application requires. A non-rejection means only that no discrepancy is detectable with the available data, so the data give no reason to correct the model.}

We propose the Fourier maximum modulus test (FMMT), which estimates the discrepancy by kernel ridge regression (KRR) \new{of the residuals} and obtains closed-form $p$-values from the joint asymptotic normality of its weighted generalized Fourier coefficients. \new{If the computer model has calibration parameters, they are estimated from the same data by the $L_2$ calibration of \citet{tuo2015efficient}, and the test accounts for this estimation (\Cref{sec:calibration}). We prove the joint asymptotic normality of all weighted Fourier coefficients, with and without estimated calibration parameters (\Cref{Th:limit,Th:cal}). The proof extends the noiseless-KRR argument of \citet{tuo2024asymptotic} from a single linear functional of KRR to a class of functionals (\Cref{Th:WeakConvergence} in the supplementary material).}

The remainder of this paper is organized as follows. \Cref{sec:formulation} formulates the testing problems and reviews related literature; \Cref{sec:methodology} presents FMMT; \Cref{sec:theory} provides theoretical justifications; \Cref{sec:calibration} extends the test to simulators with calibration parameters; \Cref{sec:simulation} reports simulation studies; \Cref{sec:case_study} presents a shear-layer application; and \Cref{sec:conc} concludes. The supplementary material contains the proofs, details of the competing tests, and additional simulation results.

\section{Problem Formulation}
\label{sec:formulation}
Let $\Omega\subset\mathbb{R}^d$ be convex and compact. Let $y^s:\Omega\rightarrow\mathbb{R}$ be the computer model output, or a surrogate model \citep{santner2003,gramacy2020surrogates} when the simulator is expensive; we treat $y^s$ as known.
The physical data $(x_i,y_i)$, $i=1,\ldots,n$, satisfy $x_i \stackrel{\text{i.i.d.}}{\sim} p(x)$ on $\Omega$ and follow the nonparametric regression model
\[y_i=f(x_i)+e_i,\]
where the \textit{physical process} $f$ is continuous and the $e_i$'s are i.i.d. random errors with zero mean and variance $\sigma^2$. In the \textit{global validation}, we test
\begin{equation}\label{GlobalTest}
    H_0: f(x)=y^s(x) \text{ for all } x\in\Omega\leftrightarrow H_1: f(x)\neq y^s(x) \text{ for some } x\in\Omega.
\end{equation}
For the \textit{subdomain validation}, we partition $\Omega$ into $N$ disjoint subdomains with positive volumes, $\Omega=\bigcup_{i=1}^N \Omega_i$, and test
\begin{equation}\label{SubdomainTest}
    H_{0i}: f(x)=y^s(x) \text{ for all } x\in\Omega_i\leftrightarrow H_{1i}: f(x)\neq y^s(x) \text{ for some } x\in\Omega_i,
\end{equation}
for $i=1,\ldots,N$, as a multiple testing problem.

Testing the equality of two regression functions $f_1$ and $f_2$, neither of which is known a priori, is a \emph{two-sample} problem; see \citet{gonzalez2013updated} for a survey and \citet{neumeyer2003nonparametric} for a residual-based test. Because $y^s$ is known, validation is instead a \emph{one-sample goodness-of-fit} problem, in which the null regression function is fully specified. \new{Examples are the order-selection test of \citet{eubank1992testing}, extended to multiple regression by \citet{aerts2000testing}, the test of \citet{fan2001goodness} based on the Fourier transform of the residuals, the generalized likelihood ratio test of \citet{fanjq2001}, and tests based on the residual-marked empirical process of \citet{stute1997nonparametric} with critical values from the Monte Carlo methods of \citet{zhu2005nonparametric}. Several of these tests have closed-form asymptotic null distributions, but resampling is commonly used with multivariate inputs or estimated parameters. FMMT obtains closed-form $p$-values for the global and all subdomain hypotheses from one joint limit, also in these cases. \Cref{sec:simulation} compares it with the tests of \citeauthor{eubank1992testing}, \citeauthor{fanjq2001}, and \citeauthor{zhu2005nonparametric}.}

\new{When the null model has parameters estimated from the same data, as in \Cref{sec:calibration}, the estimate adds to the test process a drift that is linear in the estimation error. The martingale transformation of \citet{khmaladze1981martingale} and \citet{stute1998innovation}, with a scanning family of sets for multivariate covariates \citep{khmaladze2004martingale}, removes this drift by projecting a partial-sum process along a filtration. The Fourier coefficients have no natural filtration, so we instead derive the limiting covariance of the coefficients, which accounts for the calibration estimator, and diagonalize it, for the global and the subdomain tests alike.}

Local differences have been tested with density-based \citep{duong2013local}, regression-based \citep{kim2019global}, and distance-based \citep{yan2022distance} methods. Applied methods include Gaussian process diagnostics for wind turbine power curves \citep{prakash2021gaussian} and subdomain quantile-curve tests for surface finishing \citep{jin2023hypothesis}, though these lack theoretical guarantees.

\section{Proposed Methodology}
\label{sec:methodology}
\subsection{Nonparametric regression}\label{sec:Nonpara}
\begin{newmaterial}We estimate the discrepancy $\delta:=f-y^s$ by KRR,
\begin{equation}\label{KRR:1}
    \hat{\delta}_n:=\operatorname*{argmin}_{v\in\mathcal{H}}\frac{1}{n}\sum_{i=1}^n(y_i-y^s(x_i)-v(x_i))^2+\lambda_n \|v\|^2_{\mathcal{H}},
\end{equation}\end{newmaterial}
where $\mathcal{H}$ is a reproducing kernel Hilbert space and $\lambda_n$ is the smoothing parameter. We use the isotropic Mat\'ern kernel \citep{stein1999,RasmussenWilliams06}
\begin{equation}\label{matern}
    K(x_1,x_2;\nu,\phi)=\frac{2^{1-\nu}}{\Gamma(\nu)}\left(\frac{\sqrt{2\nu}\|x_1-x_2\|}{\phi}\right)^\nu K_\nu\left(\frac{\sqrt{2\nu}\|x_1-x_2\|}{\phi}\right),
\end{equation}
where $\nu>0$ is the smoothness parameter, $\phi>0$ is the scale parameter, $K_\nu$ is the modified Bessel function of the second kind, and $\|\cdot\|$ is the Euclidean norm. By the representer theorem \citep{wahba1990spline}, $\hat{\delta}_n(x)=K(x,X)(K(X,X)+\lambda_n n I)^{-1}(Y-y^s(X))$, where $X:=(x_1,\ldots,x_n)^T$, $K(A,B):=(K(a_i,b_j;\nu,\phi))_{ij}$ for any sets of input points $A=(a_1,\ldots,a_k)^T$ and $B=(b_1,\ldots,b_l)^T$, $Y=(y_1,\ldots,y_n)^T$, and $y^s(X)=(y^s(x_1),\ldots,y^s(x_n))^T$. The choice of $\lambda_n$, $\nu$, and $\phi$ is discussed in \Cref{sec:details}.

\subsection{Global and Subdomain Tests}\label{sec:glo_sub_tests}

Let $\{h_i\}_{i=1}^\infty$ be an orthonormal basis of $L_2(\Omega)$. Each $g\in L_2(\Omega)$ has the unique representation $g=\sum_{i=1}^\infty g_i h_i$, where $g_i:=\int_\Omega g(x)h_i(x)dx$ is called a generalized Fourier coefficient, and $g=0$ if and only if all $g_i$'s are zero. We therefore consider the generalized Fourier coefficients $s^{(i)}:=\int_{\Omega} \delta(x)\sqrt{p(x)}h_i(x)dx$ of the weighted discrepancy $\delta\sqrt{p}$, where $p$ is the design density of the inputs $x_i$. If $p(x)>0$ for all $x\in\Omega$, then $H_0$ is true if and only if $s^{(i)}=0$ for all $i=1,2,\ldots$. With a density estimator $\hat{p}_n$, $s^{(i)}$ is estimated by
\begin{equation}\label{Fourier}
\hat{s}_{n}^{(i)}=\int_{\Omega} \hat{\delta}_n(x)\sqrt{\hat{p}_n(x)}h_i(x)dx.
\end{equation}
To test $s^{(i)}=0$ for all $i$ simultaneously, we use decay coefficients $\rho_i\downarrow 0$, for example $\rho_i\leq C/\log^{\ell}(i+1)$ with $\ell>1/2$ (\Cref{rem:1}). Under the conditions of \Cref{Th:limit}, $\{\sqrt{n}\rho_i\hat{s}^{(i)}_n\}_{i=1}^\infty$ converges weakly under $H_0$ to independent normal random variables with mean zero and variances $\{\rho_i^2 \sigma^2\}_{i=1}^\infty$. We therefore consider the maximum modulus test statistic
\begin{equation}\label{T}
    T:=\max_{i\geq 1}\left|\frac{\sqrt{n}\rho_i\hat{s}^{(i)}_n}{\hat{\sigma}_n}\right|=\frac{\sqrt{n}}{\hat{\sigma}_n}\max_{i\geq 1}\left|\rho_i\int_\Omega \hat{\delta}_n(x)\sqrt{\hat{p}_n(x)}h_i(x)dx\right|,
\end{equation}
where $\hat{\sigma}_n$ is a consistent estimator of $\sigma$ (\Cref{sec:details}). Under $H_0$, by \Cref{Th:limit}, Slutsky's theorem, and the continuous mapping theorem,
\begin{equation}\label{limitdist}
    \lim_{n\rightarrow\infty}\mathbb{P}(T\leq t)=\prod_{i=1}^\infty\left(2\Phi\left(\frac{t}{\rho_i}\right)-1\right)=:F(t),
\end{equation}
for each $t>0$, where $\Phi$ is the standard normal distribution function. The asymptotic $p$-value for $H_0$ is $1-F(T)$. Under $H_1$, some $s^{(i)}\neq0$, so by \Cref{rem:null}, $T$ tends to infinity at the rate $n^{1/2}$ in probability, and this \textit{Fourier maximum modulus test (FMMT)} is consistent.

The subdomain testing problem \cref{SubdomainTest} is handled similarly. For each subdomain $\Omega_i$, let $\{h_{ij}\}_{j=1}^\infty$ be an orthonormal basis of $L_2(\Omega_i)$, extended to $\Omega$ by $h_{ij}(x)=0$ for $x\in \Omega\setminus\Omega_i$. Then $H_{0i}$ is true if and only if 
$
    \int_{\Omega} \delta(x)\sqrt{p(x)}h_{ij}(x)dx=0
$
for all $j$.
Since $h_{ij}$ and $h_{st}$ have disjoint supports if $i\neq s$, $\{h_{ij}\}_{i,j=1}^\infty$ is an orthonormal basis of $L_2(\Omega)$, and \Cref{Th:limit} applies again. In analogy to \cref{T}, the test statistic for $H_{0i}$ is
\begin{equation}\label{Ti}
    T_i:=\frac{\sqrt{n}}{\hat{\sigma}_n}\max_{j\geq 1}\left|\rho_j\int_\Omega \hat{\delta}_n(x)\sqrt{\hat{p}_n(x)}h_{ij}(x)dx\right|,
\end{equation}
with $p$-value $1-F(T_i)$. We control the family-wise error rate (FWER) across the subdomains by the Bonferroni correction; \new{the procedure of \citet{holm1979simple} can also be used \citep[see][]{hochberg1987multiple}. By \Cref{Th:limit}, $T_1,\ldots,T_N$, which use the coefficients on disjoint subdomains, are asymptotically independent under $H_0$, so the procedures of \citet{vsidak1967rectangular} and \citet{hochberg1988sharper} are also asymptotically valid. Subdomains should be chosen before the data are examined, from physical knowledge of the system such as flow regimes or operating conditions.}

\subsection{Other Implementation Details}\label{sec:details}

We recommend the following implementation, which is used in \Cref{sec:simulation,sec:case_study} unless stated otherwise.
Let $\hat{f}_n$ be the KRR estimator of $f$, defined as in \cref{KRR:1} with $y_i$ in place of $y_i-y^s(x_i)$. Its smoothing parameter is $\lambda_n=C_n/n$, where $C_n$ is chosen from $\{10^{-9},10^{-8},\ldots,1\}$ by five-fold cross-validation, and the same $\lambda_n$ is used for $\hat{\delta}_n$.
The noise variance $\sigma^2$ is estimated by $\hat{\sigma}_n^2 = \sum_{i=1}^n (y_i - \hat{f}_n(x_i))^2/\{n - \text{tr}(S)\}$, which accounts for the model's complexity through the effective degrees of freedom $\text{tr}(S)$ \citep{wahba1990spline}, where 
\begin{equation}\label{S}
    S := K(X,X)(K(X,X) + \lambda_n n I)^{-1}
\end{equation}
is the smoother matrix. Such residual-based estimators are classical \citep{hall1990variance}. \new{When $y^s$ is available only at a set of simulator runs, we replace it by their kernel interpolant, computed as KRR with $C_n=10^{-10}$ (\Cref{rem:surrogate}).}

The density $p$ is estimated by a kernel density estimator \citep{silverman2018density} with product Epanechnikov kernels, one normal-reference bandwidth per coordinate, and boundary correction \citep{jones1993simple}; the estimate is normalized to integrate to one.

\new{\Cref{Th:limit} holds for any orthonormal basis. We use the Fourier basis, whose functions are ordered by frequency, which gives a natural choice of $\rho_j$, and which is easily rescaled to rectangular subdomains; the cosine basis and the Legendre polynomials share these properties.} On $[0,1]$, the orthonormal Fourier basis is $\{1, \sqrt{2}\cos(2\pi k x), \sqrt{2}\sin(2\pi k x)\}_{k=1}^\infty$; on $[0,1]^d$, tensor-product bases are used, rescaled to the side lengths of each subdomain. The coefficients \cref{Fourier} are computed by numerical integration, which also applies to irregular domains.

We set $\rho_j = 1/[\log(k_j + 2)]^\ell$ with $\ell>1/2$, where $k_j$ is the frequency of $h_j$ (in several dimensions, the sum of the directional frequencies), and order the basis so that $k_j$, and hence $\rho_j$, is monotone in $j$. Only the coefficients with $k_j\le k_{\max}$ are used, so the maximum in \cref{T} and the product in \cref{limitdist} run over these $j$. 

\begin{newmaterial}In view of the finite sample size and the truncation at $k_{\max}$, we make two additional adjustments. First, we replace the asymptotic standard deviation $\hat{\sigma}_n/\sqrt{n}$ of $\hat{s}_n^{(j)}$ in \cref{T,Ti} by an estimate of its exact standard deviation under $H_0$ given the inputs, $\hat{\sigma}_n\|w_j\|$, where $w_j:=(K(X,X)+\lambda_n n I)^{-1}\int_\Omega K(X,x)\sqrt{\hat{p}_n(x)}h_j(x)dx$ satisfies $\hat{s}_n^{(j)}=w_j^T(Y-y^s(X))$. Second, we account for the variability of $\hat{\sigma}_n$ in \cref{T,Ti}. We use the approximate $p$-value
\begin{equation*}
    1-\mathbb{E}_W\Big[\prod_{j \,:\, k_j \le k_{\max}}\Big(2\Phi\Big(\frac{TW}{\rho_j}\Big)-1\Big)\Big],
\end{equation*}
where $W:=\hat{\sigma}_n/\sigma$ is treated as independent of the coefficients. The exact distribution of $W$ depends on the noise distribution. Following \citet{satterthwaite1946approximate} and \citet[Section 4.1]{cleveland1988locally}, we use the proxy distribution by assuming the noise is normally distributed. In this case, $W^2$ approximately follows $\kappa_n\chi^2_{r_n}/r_n$ with $r_n:=\operatorname{tr}(D)^2/\operatorname{tr}(D^2)$ and $\kappa_n:=\operatorname{tr}(D)/(n-\operatorname{tr}(S))$, in which $S$ is defined in \cref{S} and $D:=(I-S)^T(I-S)$. For large sample sizes, this adjustment is negligible.\end{newmaterial}

As defaults, we use $\nu=3.5$ and $\phi=1$ with $\Omega$ rescaled to the unit cube, $\ell=0.7$, and $k_{\max}=\lfloor\sqrt{n}\rfloor$. \new{We take $\nu=3.5$, a smooth kernel, because model discrepancies are usually smooth \citep{bayarri2007framework}. We take $\ell$ close to its lower limit $1/2$ because a larger $\ell$ makes $\rho_j/\rho_1$ smaller for every $j\ge2$ and so lowers the power against discrepancies with a small coefficient $s^{(1)}$ on the constant function $h_1$. Because the KRR in \cref{KRR:1} is applied to the residuals, its regression function under $H_0$ is zero for every $\nu$. Hence $\nu$ does not affect the null distribution \cref{limitdist}, the unknown smoothness of $f$ enters the test only through $C_n$ and $\hat{\sigma}_n$, and $\nu$ matters mainly for the power. The smoothness $\nu$ can also be chosen by cross-validation together with $C_n$, which costs more computation and pays off mainly against high-frequency discrepancies.}

\section{Theoretical Justifications}\label{sec:theory}
Let $l_\infty$ denote the space of bounded real sequences, equipped with the supremum norm. Define $\hat{\delta}_n^{(i)}:=\int_\Omega(\hat{\delta}_n(x)-\delta(x))\sqrt{\hat{p}_n(x)}h_i(x)dx$.
\begin{theorem}\label{Th:limit}
    Under Conditions \ref{C:truefunction}--\ref{C:decay} below, as $n\rightarrow\infty$, the infinite-dimensional vector
\[(\sqrt{n}\rho_1\hat{\delta}_n^{(1)},\sqrt{n}\rho_2\hat{\delta}_n^{(2)},\sqrt{n}\rho_3\hat{\delta}_n^{(3)},\cdots)^T,\]
converges weakly in $l_\infty$ to a normal distribution with mean zero and covariance matrix $\operatorname{diag}(\rho_1^2\sigma^2,\rho_2^2\sigma^2,\rho_3^2\sigma^2,\cdots)$.
\end{theorem}

\begin{enumerate}[label=\rm{C}\arabic*,   
                  ref=\rm{C}\arabic*]     
    \item (Kernel and smoothness) An isotropic Mat\'ern kernel as in \cref{matern} is used, $\delta\in\mathcal{H}$.\label{C:truefunction}
    \item (Input distribution) The inputs $x_i$ are i.i.d.\ with density $p$ such that $0<\inf_{x\in\Omega}p(x)\leq \sup_{x\in\Omega}p(x)<+\infty$.\label{C:input}
    \item (Error distribution) The errors $e_i$ are i.i.d., independent of the $x_i$, and sub-Gaussian, with $\mathbb{E}e_1=0$, $\mathbb{E}e_1^2=\sigma^2\in (0,+\infty)$, and $\mathbb{E}\exp\{\vartheta e_1\}\leq \exp\{\vartheta^2\varsigma^2/2\}$ for all $\vartheta\in \mathbb{R}$ and some $\varsigma>0$.\label{C:subGaussian}
    \item (Smoothing parameter) The smoothing parameter $\lambda_n$ is non-random, $n\lambda_n=O(1)$, and $n^{2m/d}\lambda_n\rightarrow+\infty$ as $n\rightarrow\infty$, where $m:=\nu+d/2$. \label{C:tuning}
    \item (Density estimator) The density estimator satisfies $\|\sqrt{\hat{p}_n}-\sqrt{p}\|_{L_2}\xrightarrow{p}0$, and there is a set $\mathcal{Q}\subset L_\infty(\Omega)$ such that $\int_0^{+\infty}\sqrt{\log N(\epsilon,\mathcal{Q},\|\cdot\|_{L_\infty})}\,d\epsilon<\infty$ and  $\mathbb{P}(\sqrt{\hat{p}_n}\in\mathcal{Q})\rightarrow 1$ as $n\rightarrow\infty$, where, for a set of functions $\mathcal{V}$ over $\Omega$ with a semi-metric $d_\mathcal{V}(\cdot,\cdot)$, $N(\epsilon,\mathcal{V},d_\mathcal{V})$ is the smallest integer $N$ such that some functions $v_1,\ldots,v_N$ satisfy $\sup_{v\in\mathcal{V}}\min_{1\leq i\leq N}d_\mathcal{V}(v,v_i)\leq \epsilon$ \citep{VaartWCofEP00,geer2000empirical}.\label{C:density}
        \item (Decay coefficients) The decay coefficients satisfy $\int_0^{+\infty}\sqrt{\log (M(\epsilon)+1)}\,d\epsilon<\infty$, where $M(\epsilon):=\max\{i:\rho_i\geq \epsilon\}$ and we define $M(\epsilon):=0$ if no $\rho_i\geq \epsilon$.\label{C:decay}
\end{enumerate}

\begin{remark}\label{rem:1}
    Conditions \ref{C:truefunction}--\ref{C:subGaussian} are standard in the nonparametric literature \citep{geer2000empirical,van2000asymptotic}. \new{For deterministic designs, such as grids, \Cref{rem:fixed} in the supplementary material gives conditions under which \Cref{Th:limit} holds, and \Cref{sec:J-designs} reports simulations with other fixed designs.} Condition \ref{C:tuning} allows a wide range of $\lambda_n$; we recommend $\lambda_n\sim n^{-1}$, as in \Cref{sec:details}, which gives better estimation by KRR \citep{tuo2024asymptotic}.
    In Condition \ref{C:density}, $\hat{p}_n$ need not be a proper density, i.e., $\int_\Omega \hat{p}_n(x)dx\neq 1$ is allowed; if it is one, $\|\sqrt{\hat{p}_n}-\sqrt{p}\|_{L_2}$ is the Hellinger distance. Both parts of Condition \ref{C:density} hold for common density estimators, the second if the derivatives of $\hat{p}_n$ converge uniformly to those of $p$. Condition \ref{C:decay} is met if $\rho_i\leq C/\log^\ell(i+1)$ for $\ell>1/2$ with any constant $C>0$.
\end{remark}

\begin{newmaterial}
\begin{remark}\label{rem:null}
    Under $H_0$, we have $\delta=0$ and thus $\hat{\delta}_n^{(i)}=\hat{s}^{(i)}_n$, so \Cref{Th:limit} reduces to the asymptotic normality of $\sqrt{n}\rho_i\hat{s}^{(i)}_n$, which gives the null distribution \cref{limitdist}. Under $H_1$, $\hat{s}^{(i)}_n=\hat{\delta}_n^{(i)}+\int_\Omega\delta(x)\sqrt{\hat{p}_n(x)}h_i(x)dx$, where the integral converges in probability to $s^{(i)}$ by the Cauchy--Schwarz inequality and Condition \ref{C:density}, and $\hat{\delta}_n^{(i)}=O_\mathbb{P}(n^{-1/2})$ by \Cref{Th:limit}, so $\hat{s}^{(i)}_n\xrightarrow{p}s^{(i)}$ and the test is consistent.
\end{remark}

\begin{remark}\label{rem:surrogate}
    \Cref{Th:limit} treats $y^s$ as known. Let $T$ and $\hat{T}$ denote the statistic with the exact standard deviations of \Cref{sec:details}, computed with $y^s$ and with an approximation $\hat{y}^s$, such as a kernel interpolant of simulator runs. Because $\lambda_n$, $\hat{\sigma}_n$, $\hat{p}_n$, and $w_j$ do not depend on $y^s$, the Cauchy--Schwarz inequality gives
    \[
    |\hat{T}-T|\leq\rho_1\frac{\sqrt{n}\,\|\hat{y}^s-y^s\|_{X_n}}{\hat{\sigma}_n},\qquad \|v\|_{X_n}^2:=\frac{1}{n}\sum_{i=1}^n v^2(x_i).
    \]
    Hence the two tests are asymptotically equivalent whenever $\sqrt{n}\,\|\hat{y}^s-y^s\|_{X_n}\rightarrow0$ in probability, which holds for a kernel interpolant of a smooth $y^s$ when the number of runs grows fast enough (\Cref{rem:interp} in the supplementary material).
\end{remark}

\section{Validation with Calibration Parameters}\label{sec:calibration}

Computer models often contain calibration parameters with unknown values. Their estimation from physical data, known as calibration, has been studied extensively, from the Bayesian approach of \citet{kennedy2001bayesian} to the frequentist $L_2$ calibration of \citet{tuo2015efficient}; see also \citet{tuo2016theoretical}, \citet{plumlee2017bayesian}, and \citet{Wong2017}. If the parameters are estimated from data independent of the validation data, $y^s(\cdot,\hat{\theta})$ can be treated as a known function, and the tests of \Cref{sec:methodology} apply directly. Physical data are often too scarce to be split, however. When the same data are used for calibration and validation, the estimation error of $\hat{\theta}$ enters every Fourier coefficient, and the null distribution \cref{limitdist} no longer holds.

Let $y^s(\cdot,\theta)$, $\theta\in\Theta$, be the computer model, where $\Theta\subset\mathbb{R}^{d_\theta}$ is compact. Following \citet{tuo2015efficient}, we define the optimal value of the calibration parameter as the $L_2$ projection of $f$ onto the model family, and estimate it by plugging in the KRR estimator $\hat{f}_n$ for $f$:
\begin{equation}\label{eq:thetastar}
    \theta^*:=\operatorname*{argmin}_{\theta\in\Theta}\int_\Omega\big(f-y^s(\cdot,\theta)\big)^2dx ,\qquad
    \hat{\theta}:=\operatorname*{argmin}_{\theta\in\Theta}\int_\Omega\big(\hat{f}_n-y^s(\cdot,\theta)\big)^2dx .
\end{equation}
The global validation problem \cref{GlobalTest} becomes $H_0: f(x)=y^s(x,\theta^*)$ for all $x\in\Omega$ $\leftrightarrow$ $H_1: f(x)\neq y^s(x,\theta^*)$ for some $x\in\Omega$, and the subdomain hypotheses become $H_{0i}: f=y^s(\cdot,\theta^*)$ on $\Omega_i$. In this section, $\delta:=f-y^s(\cdot,\theta^*)$, so that $H_0$ states $\delta=0$. For both the global and the subdomain tests, let $\Delta\in\{\Omega,\Omega_1,\ldots,\Omega_N\}$ be the test domain, $\{h_j\}_{j=1}^\infty$ an orthonormal basis of $L_2(\Delta)$, and
\begin{equation}\label{eq:shat-cal}
    \hat{s}_j:=\int_\Delta\hat{\delta}_n^{\hat{\theta}}(x)\sqrt{\hat{p}_n(x)}h_j(x)dx ,\qquad j=1,2,\ldots ,
\end{equation}
where $\hat{\delta}_n^{\hat{\theta}}$ is the estimator in \cref{KRR:1} with $y^s(\cdot,\hat{\theta})$ in place of $y^s$.
These coefficients are correlated through $\hat{\theta}$. Under the null hypothesis, \Cref{Th:cal} in the supplementary material shows that $(\sqrt{n}\rho_1\hat{s}_1,\sqrt{n}\rho_2\hat{s}_2,\ldots)^T$ still converges weakly in $l_\infty$ to a normal distribution with mean zero, but now with the non-diagonal covariance matrix $\sigma^2\Sigma_\Delta$ given in \cref{eq:Sigma}. To obtain a statistic of the same form as \cref{T}, we diagonalize a plug-in estimate $\hat{\Sigma}_\Delta$, with eigenvalues $\hat{\tau}_k\geq0$ and orthonormal eigenvectors $\hat{u}_k=(\hat{u}_{k1},\hat{u}_{k2},\ldots)^T$, $k\geq1$, and propose
\begin{equation}\label{eq:cal-T}
    T_\Delta:=\frac{\sqrt{n}}{\hat{\sigma}_n}\sup_{k\geq1}\Big|\sum_{j\geq1}\hat{u}_{kj}\rho_j\hat{s}_j\Big|,\qquad
    p_\Delta:=1-\mathbb{E}_W\Big[\prod_{k:\,\hat{\tau}_k>0}\Big(2\Phi\Big(\frac{T_\Delta W}{\sqrt{\hat{\tau}_k}}\Big)-1\Big)\Big],
\end{equation}
where $\sqrt{\hat{\tau}_k}$ plays the role of $\rho_j$ in \cref{limitdist} and $W$ is as in \Cref{sec:details}. In practice, as in \Cref{sec:details}, only the $K_\Delta$ basis functions on $\Delta$ with frequency at most $k_{\max}$ are used, so $\hat{\Sigma}_\Delta$ is replaced by its leading $K_\Delta\times K_\Delta$ block.

We call this procedure the calibrated FMMT; \Cref{alg:cal-fmmt} in the supplementary material states it in full. As in \Cref{sec:glo_sub_tests}, $H_0$ is rejected if $p_\Omega\leq\alpha$ and $H_{0i}$ if $p_{\Omega_i}\leq\alpha/N$. Without calibration parameters, $\Sigma_\Delta=\operatorname{diag}(\rho_1^2,\rho_2^2,\ldots)$, as in \Cref{Th:limit}, and the procedure reduces to the FMMT of \Cref{sec:methodology}, except that $\hat{s}_j$ is standardized by its asymptotic standard deviation $\hat{\sigma}_n/\sqrt{n}$.

\section{Simulation Studies}\label{sec:simulation}

\subsection{Simulation design}\label{subsec:design}

The data are generated from $y_i=y^s(x_i)+c\,\delta_0(x_i)+e_i$, $i=1,\ldots,n$, where $e_i\sim N(0,\sigma^2)$ with $\sigma=0.5$. In one dimension, $\Omega=[0,1]$, $y^s(x)=e^x$, $n=25$ or $50$, and the inputs are drawn from the $N(0.5,0.2^2)$ distribution truncated to $[0,1]$. We use six discrepancy shapes $\delta_0$: a linear trend, a sine wave, a narrow bump, a wide bump, a step, and a boundary layer; their formulas are given in \Cref{sec:G-settings} of the supplementary material. We also consider the simple machine of \citet[Section 3.1]{brynjarsdottir2014learning}, in which the model $y^s(x)=0.65x$ ignores a frictional loss and the physical process is $0.65x/(1+cx)$; we draw the inputs uniformly on $[0,4]$. In two dimensions, $\Omega=[0,1]^2$, the inputs are uniform, $n=200$, and $\delta_0$ is a bump, a step, or a product of sine waves.

We normalize $\delta_0$ so that $\|\delta_0\|_{L_2(p)}=1$ and take the strength $c=2.80\,\sigma/\sqrt{n}$; then the two-sided $z$-test that knows $\delta_0$ (the oracle test) has power 0.80 at $\alpha=0.05$. For the simple machine, $c$ is chosen so that this oracle test has power 0.80. All discrepancies are therefore equally hard to detect for a test that knows their shape, and none is tuned to a particular test.

We compare FMMT with the order-selection test of \citet{eubank1992testing} (EH), the generalized likelihood ratio test of \citet{fanjq2001} (GLR), the nonparametric Monte Carlo test of \citet{zhu2005nonparametric} (NMCT), and the two-sample test of \citet{neumeyer2003nonparametric} (ND); EH and ND are used only in one dimension, for which they were proposed. EH, GLR, and NMCT are applied to the residuals $y_i-y^s(x_i)$, and EH includes the constant in the cosine basis because $H_0$ has no unknown parameter. ND compares the physical data with $n$ simulator outputs at independently drawn inputs. Details are in \Cref{sec:G-competitors} of the supplementary material.

Each setting is replicated 1000 times, so a rejection rate near 0.05 has Monte Carlo standard error about 0.007, and Type~I error rates between 0.037 and 0.063 agree with the nominal level. We compare power only among tests whose rates do not exceed 0.063.

\subsection{Global tests}\label{subsec:global}

FMMT holds the nominal level in all five settings, with Type~I error rates between 0.047 and 0.055 (\Cref{fig:comparison}(a)), and so do GLR, with rates between 0.040 and 0.059, and NMCT, except in two dimensions (0.064). EH exceeds 0.063 in all four one-dimensional settings (0.065 to 0.075), so its power, although the highest in 12 of the 14 one-dimensional cases, is not comparable. At $\alpha=0.10$, the power of FMMT is between 0.03 below and 0.11 above that of EH at $\alpha=0.05$ (\Cref{sec:G-rates}), so EH remains a strong competitor in one dimension. ND exceeds 0.063 for the simple machine (0.076 and 0.083).

Among the remaining tests, FMMT and GLR are the most powerful in one dimension (\Cref{fig:comparison}(b) and (c)). FMMT has higher power for the wide bump and the simple machine, GLR for the narrow bump, and the two differ by less than 0.05 for the other four shapes. NMCT and ND are less powerful than FMMT in every one-dimensional case. In two dimensions, FMMT has the highest power for all three shapes, 0.20, 0.33, and 0.22, compared with 0.12 to 0.13 for GLR (\Cref{fig:comparison}(d)).

\begin{figure}[!t]
    \centering
    \includegraphics[width=0.95\textwidth]{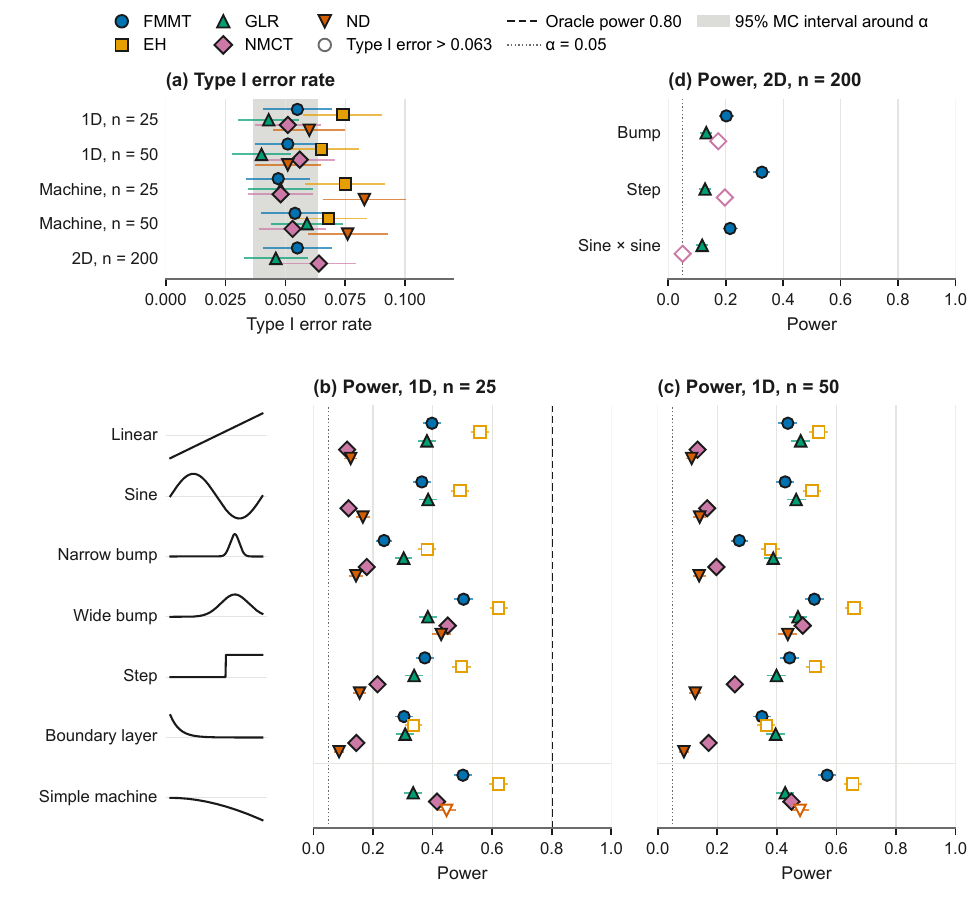}
    \captionsetup{font={color=revcolor}}\caption{Rejection rates at $\alpha=0.05$ over 1000 replications, with $\pm2$ Monte Carlo standard errors. (a)~Type~I error rates, with the 95\% Monte Carlo interval around $\alpha$ shaded. (b)--(d)~Power; open symbols mark tests whose Type~I error rate lies above this interval.}
    \label{fig:comparison}
\end{figure}

\subsection{Subdomain tests}\label{subsec:subdomain}

We illustrate the subdomain tests in two dimensions, with the design of \Cref{subsec:design}, $n=200$, and $\Omega=[0,1]^2$ divided into four quadrants of side 1/2. The discrepancy is $\sin(2\pi x_1)\sin(2\pi x_2)$ on the quadrant $[0.5,1]^2$ and zero elsewhere, with strength $c$ or $2c$. \Cref{tab:subdomain} reports the rejection rates of the global test and of the subdomain test on $[0.5,1]^2$; under $H_0$, the latter is the FWER of the four subdomain tests.

The subdomain test detects the local discrepancy more often than the global test, and in about a fifth of the replications only the subdomain test rejects, so the subdomain tests add information to the global test. Under $H_0$, when both families are tested at level 0.05, at least one rejects in 7.2\% of the replications; testing each family at level $\alpha/2$ keeps this rate at most $\alpha$. The procedures of \citet{holm1979simple}, \citet{hochberg1988sharper}, and \citet{vsidak1967rectangular} raise the power of the subdomain test by at most 0.01 over the Bonferroni correction. At least one of the three subdomains without the discrepancy is rejected in 4.5\% of the replications at $c$ and 9.0\% at $2c$, because the KRR estimate uses the data on both sides of a subdomain boundary; rejections on adjacent subdomains should therefore be read together. \Cref{sec:H} of the supplementary material reports other partitions and the one-dimensional case.

\begin{table}[!t]
\centering
\captionsetup{font={color=revcolor}}\caption{Rejection rates in the two-dimensional subdomain example ($2\times2$ partition, Bonferroni correction, 1000 replications independent of those in \Cref{fig:comparison}). ``Subdomain only'': the test in the column ``Subdomain test'' rejects and the global test does not. ``Either'': the global test or any of the four subdomain tests rejects.}
\label{tab:subdomain}
\begin{tabular}{lcccc}
\toprule
 & Global test & Subdomain test & Subdomain only & Either \\
\midrule
No discrepancy ($H_0$) & 0.051 & 0.044 & 0.021 & 0.072 \\
Quadrant, strength $c$ & 0.23 & 0.35 & 0.18 & 0.43 \\
Quadrant, strength $2c$ & 0.77 & 0.95 & 0.19 & 0.96 \\
\bottomrule
\end{tabular}
\end{table}

\subsection{Tests with an unknown calibration parameter}\label{subsec:calibration}

We now estimate the calibration parameter as in \Cref{sec:calibration}. The first example is the simple machine with $\theta$ unknown: $y^s(x,\theta)=\theta x$ on $[0,4]$, and the data come from $0.65x$ under $H_0$ and from $0.65x/(1+cx)$ under the alternative. Because calibration absorbs most of the frictional loss, $c$ is chosen as in \Cref{subsec:design} for the remainder (\Cref{sec:I-settings} of the supplementary material), and $\sigma=0.1$. The second example, the exponential model $y^s(x,\theta)=\theta_1\exp(\theta_2x)$ with $\theta=(1,0.8)$, is nonlinear in $\theta$. Its inputs on $[0,1]$ have density $0.5+x$, and $\sigma=0.1$. The discrepancy is a smooth bump with $L_2$ norm 0.3 on $[2/3,1]$, orthogonal to the gradients of $y^s$, so that $\theta^*$ equals the true value.

We compare the calibrated FMMT with the naive FMMT, which substitutes $\hat{\theta}$ and ignores its estimation error, with GLR, which re-estimates $\theta$ in every bootstrap draw, and with NMCT, which accounts for the estimation by a linearization (\Cref{sec:I-settings} of the supplementary material). \Cref{tab:calibration} reports the rejection rates of the global tests for $n=50$ and 100.

GLR and NMCT hold the nominal level, and the calibrated FMMT is slightly conservative, with Type~I error rates between 0.032 and 0.044. FMMT is more powerful than GLR in all four cases. For the local discrepancy of the exponential model, it is also more powerful than NMCT, with more than twice its power at $n=50$; for the frictional loss, NMCT is the most powerful. The naive FMMT is too conservative and, for the simple machine, has about half the power of the calibrated FMMT, so the covariance correction of \Cref{sec:calibration} is needed. When the discrepancy lies on one of three subdomains, the calibrated subdomain test detects it more often than the global test, but through $\hat{\theta}$ it also raises the family-wise error rate on the other two subdomains, to at most 0.14 with $n=100$ and 0.07 with $n=200$ (\Cref{sec:I-interaction} of the supplementary material).

\begin{table}[!t]
\centering
\captionsetup{font={color=revcolor}}\caption{Rejection rates of the global tests with an unknown calibration parameter at $\alpha=0.05$ over 1000 replications. ``Naive FMMT'': FMMT with $\hat{\theta}$ substituted and no covariance correction.}
\label{tab:calibration}
\begin{tabular}{lcccccccc}
\toprule
 & \multicolumn{4}{c}{Simple machine} & \multicolumn{4}{c}{Exponential} \\
\cmidrule(lr){2-5}\cmidrule(lr){6-9}
 & \multicolumn{2}{c}{$H_0$} & \multicolumn{2}{c}{Frictional loss} & \multicolumn{2}{c}{$H_0$} & \multicolumn{2}{c}{Local bump} \\
\cmidrule(lr){2-3}\cmidrule(lr){4-5}\cmidrule(lr){6-7}\cmidrule(lr){8-9}
Test & $n=50$ & 100 & 50 & 100 & 50 & 100 & 50 & 100 \\
\midrule
Calibrated FMMT & 0.041 & 0.044 & 0.51 & 0.54 & 0.032 & 0.040 & 0.78 & 0.97 \\
GLR & 0.047 & 0.056 & 0.45 & 0.50 & 0.053 & 0.060 & 0.60 & 0.94 \\
NMCT & 0.049 & 0.052 & 0.62 & 0.65 & 0.049 & 0.059 & 0.35 & 0.84 \\
\midrule
Naive FMMT & 0.014 & 0.012 & 0.25 & 0.23 & 0.020 & 0.018 & 0.76 & 1.00 \\
\bottomrule
\end{tabular}
\end{table}

\subsection{Other settings}\label{subsec:other}

\Cref{sec:J} of the supplementary material reports the sensitivity of FMMT in the one-dimensional setting of \Cref{subsec:design} with $n=50$. When $\nu$ varies from 1.5 to 5.5, $\phi$ from 0.5 to 2, $\ell$ from 0.6 to 1.5, or $k_{\max}$ from 3 to 14, the Type~I error rate stays between 0.042 and 0.058. The power changes by at most 0.03, except that a larger $\ell$ lowers it by up to 0.25, most for the linear trend and the sine wave, and a smaller $\nu$ raises it against high-frequency discrepancies. Choosing $\nu$ by cross-validation keeps the level (0.054) and raises the power against $\cos(12\pi x)$ at strength $2c$ from 0.07 to 0.31. With cosine, Legendre, wavelet, and kernel eigenfunction bases, the Type~I error rate is between 0.047 and 0.049, and no basis is the most powerful for every shape.

For grid, Latin hypercube, Sobol$'$, and Halton designs in one and two dimensions, redrawn in each replication or fixed, the Type~I error rate of FMMT lies between 0.035 and 0.065, at most 0.002 outside the interval from 0.037 to 0.063. Without the adjustment for the estimated $\sigma$ in \Cref{sec:details}, the Type~I error rates in the null settings of \Cref{subsec:design} are 0.069 to 0.089 in one dimension and 0.060 in two dimensions.
\end{newmaterial}

\section{Application to Shear Layer Experiment}\label{sec:case_study}

We apply FMMT to the compressible shear layer, a benchmark for turbulence models \citep{oberkampf2006measures,wang2009bayesian}, with the convective Mach number $M_c$ as input and the compressibility factor as output. The data, compiled by \citet{wang2009bayesian}, consist of 11 simulator runs at equally spaced values of $M_c$ from 0.1 to 1.5 and $n=32$ physical observations \citep{oberkampf2006measures}. \new{We replace $y^s$ by the kernel interpolant of the runs (\Cref{sec:details}) and test on $\Omega=[0,1.5]$. Without physical knowledge of where the model may fail, we divide $\Omega$ into six subdomains of equal length 0.25, which contain 2, 5, 12, 9, 2, and 2 observations.}

\begin{figure}[!t]
    \centering
    \includegraphics[width=0.92\textwidth]{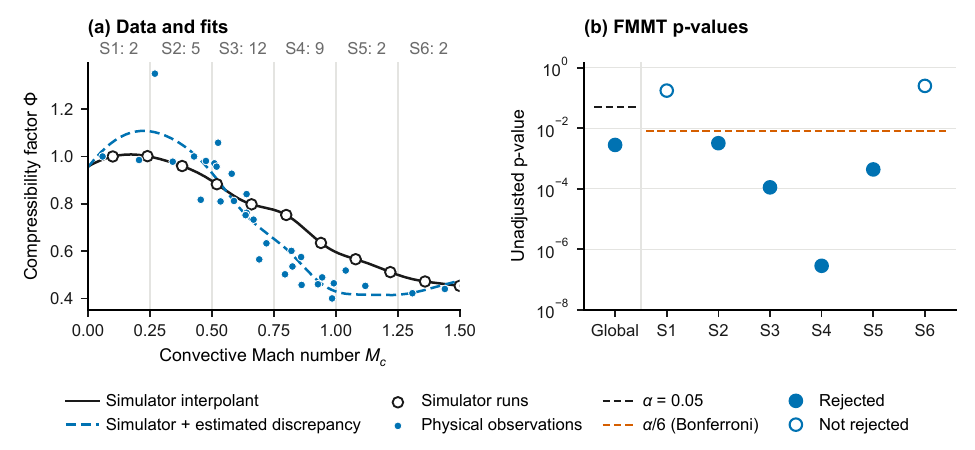}
    \caption{Compressible shear layer. (a)~Simulator runs and their kernel interpolant, the physical observations, and the interpolant plus the estimated discrepancy; the numbers at the top give the observations in each subdomain. (b)~Unadjusted $p$-values, with the levels $\alpha=0.05$ and $\alpha/6$; filled symbols mark rejections.}
    \label{fig:shear_layer_viz}
\end{figure}

The global test rejects $H_0$ with $p=0.0028$, whereas the two-sample test of \citet{neumeyer2003nonparametric} does not reject ($p=0.165$ for $K_N^{(1)}$ and $0.29$ for $K_N^{(2)}$). With the Bonferroni correction, the subdomain tests reject on subdomains 2 to 5 ($0.25\le M_c<1.25$) but not on subdomains 1 and 6\new{, where two observations give the tests little power} (\Cref{fig:shear_layer_viz}(b)). \new{With level $\alpha/2$ for each family, as in \Cref{subsec:subdomain}, the conclusions are unchanged.}

\new{Because $n$ is small and $y^s$ is interpolated, we check these conclusions in \Cref{sec:K} of the supplementary material. If the simulator were a smooth logistic curve fitted to the runs, replacing it by its interpolant would change the test statistic by at most 0.09 (\Cref{rem:surrogate}), against the observed 4.32, and data generated from this simulator at the 32 observed inputs give Type~I error rates of 0.048 for the global test and 0.034 to 0.044 for the unadjusted subdomain tests. On the physical data, a parametric bootstrap and the 11 interpolants that each leave out one run give the same rejections.}

\new{\citet{oberkampf2006measures} found by nonlinear regression that the model underestimates the compressibility factor for $0.3\le M_c\le0.6$ and overestimates it for $0.7\le M_c\le1.3$, as does the estimated discrepancy, which is positive for $M_c$ below about 0.55 and negative between about 0.55 and 1.45 (\Cref{fig:shear_layer_viz}(a)).} \citet{wang2009bayesian} rejected the model at $M_c=0.8$, in subdomain 4, as FMMT does. \new{At $M_c=0.65$, in subdomain 3, their interval for the discrepancy excludes zero but lies within their tolerance of 0.12, so they accepted the model. Our rejection there thus reveals a discrepancy that is small by their criterion.} At $M_c=1.2$, in subdomain 5, they found the data too sparse to decide. \new{Our rejection on subdomain 5 also rests on only two observations and partly reflects the large discrepancy in subdomain 4, because the KRR estimate uses the data on both sides of a subdomain boundary; it should be interpreted with caution.}

\section{Concluding Remarks}
\label{sec:conc}

This work introduces a hypothesis testing methodology for validating computer models globally or locally. \new{The joint asymptotic normality of the weighted generalized Fourier coefficients, with and without estimated calibration parameters, gives closed-form asymptotic $p$-values for both the global and the subdomain tests. The behavior of the test with more than two inputs remains to be studied.} A natural extension is to test whether $f$ and $y^s$ agree up to a prespecified tolerance, since small deviations may be acceptable for the intended application. \new{Another is to allocate the physical observations to the subdomains of interest.}

\section{Data and Code Availability Statement}

The shear-layer data used in \Cref{sec:case_study} are available at
\href{https://doi.org/10.1198/TECH.2009.07011}{doi:10.1198/TECH.2009.07011}. All code to reproduce the simulation studies and the shear-layer application is publicly available at \href{https://github.com/ChaoanLi/FMMT}{https://github.com/ChaoanLi/FMMT}.

\bibliography{bibliography.bib}

\newpage
\phantomsection\label{supplementary-material}
\bigskip

\begin{center}

{\large\bf SUPPLEMENTARY MATERIAL}

\end{center}

\begin{description}
\item[Supplementary Appendix:]
Proofs of \Cref{Th:limit}, including a general weak convergence theorem for KRR-based inner-product processes (\Cref{sec:A,sec:B}); the covariance, algorithm, and asymptotic theory of the calibrated test of \Cref{sec:calibration}, with proofs (\Cref{sec:C,sec:D}); conditions for deterministic designs and for simulators known only at a set of runs (\Cref{rem:fixed,rem:interp}); and details and additional results of the simulation studies and the shear-layer application (\Cref{sec:G,sec:H,sec:I,sec:J,sec:K}). (PDF file)
\end{description}

\begin{bibunit}
\appendix\label{sec:app}
\counterwithin{table}{section}
\counterwithin{figure}{section}

\section{A general weak convergence theorem}\label{sec:A}

Throughout \Cref{sec:A,sec:B}, $y^s$ is a fixed function and $\hat{\delta}_n$ is the estimator in \cref{KRR:1}. With $y^s\equiv0$, we have $\delta=f$ and $\hat{\delta}_n=\hat{f}_n$, the KRR estimator of $f$ in \Cref{sec:details}; \Cref{sec:D} uses this case.

The proof of \Cref{Th:limit} relies on the following result on the weak convergence of stochastic processes defined over a set of functions $\mathcal{G}\subset L_2(\Omega)$:
\[Z_n(g):=\sqrt{n}\int_\Omega (\hat{\delta}_n-\delta)(x)g(x)dx,\quad\quad g\in\mathcal{G}.\]

\begin{theorem}\label{Th:WeakConvergence}
    Suppose Conditions \ref{C:truefunction}--\ref{C:tuning} hold, and that the function space $\mathcal{G}$ satisfies the entropy integral condition
    \begin{equation}\label{entropyintegral}
        \int_0^\infty \sqrt{\log N(\epsilon,\mathcal{G},\|\cdot\|_{L_2})}d\epsilon <\infty.
    \end{equation}
    Then we have
    \begin{equation}\label{ZnConvergence}
        Z_n\xrightarrow{\mathscr{L}}GP(0,\Psi), \text{ in } L_\infty(\mathcal{G}),
    \end{equation}
with $\Psi(g_1,g_2)=\sigma^2\int_\Omega g_1(x)g_2(x)/p(x)dx$.
\end{theorem}

With \Cref{Th:WeakConvergence}, we can prove \Cref{Th:limit}.

\begin{proof}[Proof of \Cref{Th:limit}]
       Let $\mathcal{G}:=\{\rho_i h_i q:i\in\mathbb{N}_+,q\in\mathcal{Q}\}$. We first verify that $\mathcal{G}$ satisfies condition \cref{entropyintegral}. Let $\mathcal{G}_i:=\{\rho_i h_i q:q\in\mathcal{Q}\}$. Without loss of generality, assume $\rho_i\leq 1$ for all $i$. To bound $N(\epsilon,\mathcal{G}_i,\|\cdot\|_{L_2})$, we note that
    \[\|\rho_i h_i q_1-\rho_i h_i q_2\|_{L_2}\leq \rho_i \|q_1-q_2\|_{L_\infty}\leq \|q_1-q_2\|_{L_\infty}.\]
    This implies
    \[N(\epsilon,\mathcal{G}_i,\|\cdot\|_{L_2})\leq  N(\epsilon,\mathcal{Q},\|\cdot\|_{L_\infty}).\]
    On the other hand, Condition \ref{C:density} implies that $\mathcal{Q}$ is bounded in $L_\infty(\Omega)$. Thus we have
    \[\|\rho_i h_i q\|_{L_2}\leq \rho_i \|q\|_{L_\infty}\leq C\rho_i\]
    for some $C>0$. In other words, whenever $C\rho_i<\epsilon$, that is, $i>M(\epsilon/C)$, we only need one $\epsilon$-ball to cover all $\mathcal{G}_i$. Hence, we obtain
    \begin{eqnarray*}
        N(\epsilon,\mathcal{G},\|\cdot\|_{L_2})\leq M(\epsilon/C)N(\epsilon,\mathcal{Q},\|\cdot\|_{L_\infty})+1\leq (M(\epsilon/C)+1)N(\epsilon,\mathcal{Q},\|\cdot\|_{L_\infty}).
    \end{eqnarray*}
    So we have
    \begin{eqnarray*}
        &&\int_0^{+\infty} \sqrt{\log N(\epsilon,\mathcal{G},\|\cdot\|_{L_2})} d\epsilon
        \leq \int_0^{+\infty} \sqrt{\log (M(\epsilon/C)+1)+\log N(\epsilon,\mathcal{Q},\|\cdot\|_{L_\infty})} d\epsilon\\
        &\leq& \int_0^{+\infty} \sqrt{\log (M(\epsilon/C)+1)} d\epsilon+\int_0^{+\infty} \sqrt{\log N(\epsilon,\mathcal{Q},\|\cdot\|_{L_\infty})} d\epsilon\\
        &=& C\int_0^{+\infty} \sqrt{\log (M(\epsilon)+1)} d\epsilon+\int_0^{+\infty} \sqrt{\log N(\epsilon,\mathcal{Q},\|\cdot\|_{L_\infty})} d\epsilon<\infty.
    \end{eqnarray*}
    We can now invoke \Cref{Th:WeakConvergence} to conclude that $Z_n(i,q):=\sqrt{n}\int_\Omega (\hat{\delta}_n-\delta)(x)\rho_iq(x)h_i(x)dx$ with $i\in\mathbb{N}_+,q\in\mathcal{Q}$ converges weakly to a Gaussian process $Z$. According to Addendum 1.5.8 of \citet{VaartWCofEP00}, $Z(i,\cdot)$ has $L_2(\Omega)$-continuous sample paths for each $i$. Therefore, given the conditions that $\mathbb{P}(\sqrt{\hat{p}_n}\in\mathcal{Q})\rightarrow 1$ and $\|\sqrt{\hat{p}_n}-\sqrt{p}\|_{L_2}\xrightarrow{p}0$, we can apply the continuous mapping theorem to obtain that $Z_n(\cdot,\sqrt{\hat{p}_n})$ converges weakly to the infinite-dimensional random vector $(\rho_1\sigma S_1,\rho_2\sigma S_2,\ldots)$ with $S_1,S_2,\ldots$ being independent and identically distributed $N(0,1)$ random variables. This proves the desired result by employing the continuous mapping theorem once again.
\end{proof}

\section{Proof of \Cref{Th:WeakConvergence}}\label{sec:B}
In view of the continuity of the $L_2$ inner product, $Z_n$ has continuous sample paths, and thus is separable. In other words, $\mathcal{G}$ can essentially be regarded as a countable set, which effectively avoids measurability issues.

To prove that $Z_n$ converges weakly to a Gaussian process, according to the classic theory on weak convergence of stochastic processes (see, for example, Theorems 1.5.4 and 1.5.7 of \citet{VaartWCofEP00}), it suffices to prove the following three conditions:
\begin{enumerate}[label=\rm{B}\arabic*,   
                  ref=\rm{B}\arabic*]     
    \item Total boundedness: $N(\epsilon,\mathcal{G},\|\cdot\|_{L_2})<+\infty$ for each $\epsilon>0$.\label{B1_bound}
    \item Asymptotic equicontinuity: for every $\epsilon,\eta>0$, there exists $\kappa>0$ such that
    \[\limsup_{n\rightarrow\infty}\mathbb{P}\left(\sup_{\stackrel{g_1,g_2\in\mathcal{G}}{\|g_1-g_2\|_{L_2}<\kappa}}|Z_n(g_1)-Z_n(g_2)|>\epsilon\right)<\eta.\]\label{B2_asymp}
    \item Each finite-dimensional distribution of $Z_n$ converges weakly to the corresponding finite-dimensional distribution of $GP(0,\Psi)$.\label{B3_finite}
\end{enumerate}

Among the above three conditions, Condition \ref{B1_bound} is trivially true in view of the entropy integral condition \cref{entropyintegral}. We shall prove Conditions \ref{B2_asymp} and \ref{B3_finite} in the following two subsections.

\subsection{Asymptotic equicontinuity}\label{Sec:AsyEqui}

We shall prove the asymptotic equicontinuity using the device introduced by \citet{tuo2024asymptotic}. The first step is to partition $Z_n$ into a bias term and a variance term. As $Z_n$ is a stochastic process, we call the two terms the \textit{bias process} and the \textit{variance process}, denoted as $B_n(g)$ and $V_n(g)$, respectively. They are defined as
\begin{eqnarray*}
    B_n(g)&:=&\mathbb{E}[Z_n(g)|X_n],\\
    V_n(g)&:=&Z_n(g)-\mathbb{E}[Z_n(g)|X_n],
\end{eqnarray*}
where $X_n:=(x_1,\ldots,x_n)$. Clearly, to prove Condition \ref{B2_asymp}, it suffices to prove that both $B_n$ and $V_n$ are asymptotically equicontinous, i.e.,
\begin{equation}\label{equicontinuity_B} \limsup_{n\rightarrow\infty}\mathbb{P}\left(\sup_{\stackrel{g_1,g_2\in\mathcal{G}}{\|g_1-g_2\|_{L_2}<\kappa}}|B_n(g_1)-B_n(g_2)|>\epsilon\right)<\eta.
\end{equation}
and
\begin{equation}\label{equicontinuity_V}
\limsup_{n\rightarrow\infty}\mathbb{P}\left(\sup_{\stackrel{g_1,g_2\in\mathcal{G}}{\|g_1-g_2\|_{L_2}<\kappa}}|V_n(g_1)-V_n(g_2)|>\epsilon\right)<\eta.
\end{equation}

\subsubsection{Useful results from \citet{tuo2024asymptotic}}
A major auxiliary function introduced in \citet{tuo2024asymptotic} is called the \textit{noiseless KRR}. 
Given any $W_n=(w_1,\ldots,w_n)\in\Omega^n$, define the $(W_n,2)$-norm as $\|v\|_{W_n}^2:=n^{-1}\sum_{i=1}^n v^2(w_i)$ for any function $v$. Let $v_g:=\int_\Omega K(\cdot,x)g(x)dx$. To align with the current context, we define the noiseless KRR as a mapping $v_g\mapsto \hat{v}_{g,W_n}$ given $W_n\in\Omega^n$, with
\[\hat{v}_{g,W_n}:=\operatorname*{argmin}_{v\in\mathcal{H}}\|v-v_g\|_{W_n}^2+\lambda_n \|v\|^2_{\mathcal{H}}.\]
It is worth noting that, if $W_n$ is deterministic, so is $\hat{v}_{g,W_n}$.

As shown in \citet{tuo2024asymptotic}, the properties of noiseless KRR are deeply connected with those of $B_n$ and $V_n$.  Below is a list of important properties proved in \citet{tuo2024asymptotic} under Conditions 
\ref{C:truefunction}--\ref{C:input}.
\begin{enumerate}[label=\rm{P}\arabic*,   
                  ref=\rm{P}\arabic*]     
    \item  $|B_n(g)|=\sqrt{n}\left|\langle \delta,\hat{v}_g-v_g\rangle_\mathcal{H}\right|\leq \sqrt{n}\|\hat{v}_g-v_g\|_\mathcal{H}\|\delta\|_\mathcal{H}$.\label{P1}
    \item $V_n(g)=\sqrt{n}\sum_{i=1}^n u_i(g)e_i$, with $u_i(g)=\lambda^{-1}n^{-1}(v_g(x_i)-\hat{v}_g(x_i))$.\label{P2}
    \item Suppose $n^{-m/d}\lambda^{-1/2}$ is bounded. Then for each $\eta>0$, there exists $C_\eta>0$ independent of $n$ and $\lambda$, such that the following set $E_n(\eta)\subset \Omega^n$ satisfies $\mathbb{P}(X_n\in E_n(\eta))>1-\eta$:
    \begin{eqnarray*}
        E_n(\eta):=\Big\{W_n\in\Omega^n:\|\hat{v}_{g,W_n}-v_g\|_{W_n}\leq C_\eta \lambda\|g\|_{L_2},\\
        \|\hat{v}_{g,{W_n}}-v_g\|_\mathcal{H}\leq C_\eta \lambda^{\frac{1}{2}}\|g\|_{L_2}, \text{ for all }  g\in L_2\Big\}.
    \end{eqnarray*}\label{P3}
\end{enumerate}

\subsubsection{Bias term}
First we prove \cref{equicontinuity_B}. Because $B_n$ is linear, \cref{equicontinuity_B} is equivalent to
\begin{equation}\label{equicontinuity_B1} \limsup_{n\rightarrow\infty}\mathbb{P}\left(\sup_{\stackrel{g\in\mathcal{G}}{\|g\|_{L_2}<\kappa}}|B_n(g)|>\epsilon\right)<\eta.
\end{equation}
We now invoke Properties \ref{P1} and \ref{P3} to obtain
\[\mathbb{P}\left(\sup_{\|g\|_{L_2}<\kappa}|B_n(g)|\leq \sqrt{n}C_\eta \lambda^{1/2}\kappa\|\delta\|_\mathcal{H}\right)>1-\eta,\]
which, together with the condition $n\lambda_n=O(1)$ from Condition \ref{C:tuning}, leads to \cref{equicontinuity_B1}. This also proves \cref{equicontinuity_B}.

\subsubsection{Variance term}
Next, we consider \cref{equicontinuity_V}. Its proof follows from a similar argument as that for Theorem S2.3 of \citet{tuo2024asymptotic}. As a version of Dudley's theorem, Corollary 2.2.8 of \citet{VaartWCofEP00} states that a zero-mean sub-Gaussian process with respect to a semi-metric $d_Z$, i.e., a stochastic process $Z(g)$ satisfying $\mathbb{E}\exp\{\vartheta(Z(g_1)-Z(g_2))\}\leq \exp\{\vartheta^2 d^2_Z(g_1,g_2)/2\}$ for all possible $\vartheta,g_1,g_2$, is subject to the following uniform bound
\[\mathbb{E}\left[\sup_{\stackrel{g_1,g_2\in\mathcal{G}}{d_Z(g_1,g_2)<\kappa}}|Z(g_1)-Z(g_2)|\right]\leq A \int_0^\kappa \sqrt{\log N(\epsilon, \mathcal{G},d_Z)}d\epsilon,\]
for a universal constant $A$.

To make use of the above maximum inequality for sub-Gaussian processes, we need to suppress the input location $X_n$ and consider probabilities with respect to $e_i$'s alone. This is legitimate
because in view of the independence between $X_n$ and $e_i$'s, we have a product probability space on which we can apply Fubini's Theorem. Consider the set $E_n(\eta/2)$ defined in Property \ref{P3} and let $X_0\in E_n(\eta/2)$.
We now obtain
\begin{eqnarray*}
   &&\mathbb{E}\left[\exp\{\vartheta (V_n(g_1)-V_n(g_2))\}|X_n=X_0\right]\\
   &=&\mathbb{E}\left[\exp\{\vartheta V_n(g_1-g_2)\}|X_n=X_0\right]\\
   &=&\prod_{i=1}^n\mathbb{E}[\exp\{\vartheta\sqrt{n}u_i(g_1-g_2)e_i\}|X_n=X_0]\\
   &\leq& \exp\left\{\vartheta^2\varsigma^2n\sum_{i=1}^n u_i^2(g_1-g_2)/2\right\} \\
   &=&\exp\left\{\vartheta^2\varsigma^2{n}\lambda^{-2}n^{-2}\sum_{i=1}^n (\hat{v}_{g_1-g_2,X_0}(x_i)-v_{g_1-g_2}(x_i))^2/2\right\}\\
   &=&\exp\left\{\vartheta^2\varsigma^2\lambda^{-2}\|\hat{v}_{g_1-g_2,X_0}-v_{g_1-g_2}\|^2_{X_0}/2\right\}\\
   &\leq &\exp\left\{\vartheta^2\varsigma^2\lambda^{-2}C_{\eta/2}^2\lambda^2\|g_1-g_2\|_{L_2}^2/2\right\}=\exp\left\{\vartheta^2\varsigma^2C_{\eta/2}^2\|g_1-g_2\|_{L_2}^2/2\right\},
\end{eqnarray*}
where the first equality follows from the fact that $V_n$ is linear; the second equality follows from the notation in Property \ref{P2}; the first inequality follows from the condition $\mathbb{E}\exp\{\vartheta e_1\}\leq \exp\{\vartheta^2\varsigma^2/2\}$; the third equality follows from Property \ref{P2}; and the last inequality follows from Property \ref{P3}. Therefore, we conclude that when $X_n=X_0$, $V_n(g)$ is a sub-Gaussian process with respect to the semi-norm 
$d_{V_n}(g_1,g_2):=\varsigma C_{\eta/2} \|g_1-g_2\|_{L_2}.$
This enables us to use Dudley's inequality to assert
\begin{eqnarray*}
    &&\mathbb{E}\left[\sup_{\stackrel{g_1,g_2\in\mathcal{G}}{\|g_1-g_2\|_{L_2}<\kappa}}\left|V_n(g_1)-V_n(g_2)\right|\Bigg|X_n=X_0\right]\\
    &\leq& A \int_0^{\varsigma C_{\eta/2}\kappa} \sqrt{\log N(\epsilon, \mathcal{G},d_{V_n})}d\epsilon\\
    &=&A \int_0^{\varsigma C_{\eta/2}\kappa} \sqrt{\log N(\varsigma^{-1} C^{-1}_{\eta/2}\epsilon, \mathcal{G},\|\cdot\|_{L_2})}d\epsilon\\
    &=&A\varsigma C_{\eta/2}\int_0^{\kappa} \sqrt{\log N(\epsilon, \mathcal{G},\|\cdot\|_{L_2})}d\epsilon
\end{eqnarray*}
Because $\int_0^\infty \sqrt{\log N(\epsilon, \mathcal{G},\|\cdot\|_{L_2})}d\epsilon<\infty$, we can choose $\kappa$ sufficiently small such that 
\begin{eqnarray*}
&&\mathbb{P}\left(\sup_{\stackrel{g_1,g_2\in\mathcal{G}}{\|g_1-g_2\|_{L_2}<\kappa}}|V_n(g_1)-V_n(g_2)|>\epsilon\Bigg|X_n=X_0\right)\\
&\leq&
    \mathbb{E}\left[\sup_{\stackrel{g_1,g_2\in\mathcal{G}}{\|g_1-g_2\|_{L_2}<\kappa}}\left|V_n(g_1)-V_n(g_2)\right|\Bigg|X_n=X_0\right]/\epsilon<\eta/2,
\end{eqnarray*}
where we use Markov's inequality to establish the first inequality. Finally, we use the total law of probability to arrive at
\begin{eqnarray*}
    &&\mathbb{P}\left(\sup_{\stackrel{g_1,g_2\in\mathcal{G}}{\|g_1-g_2\|_{L_2}<\kappa}}|V_n(g_1)-V_n(g_2)|>\epsilon\right)\\
    &=&\mathbb{P}\left(\sup_{\stackrel{g_1,g_2\in\mathcal{G}}{\|g_1-g_2\|_{L_2}<\kappa}}|V_n(g_1)-V_n(g_2)|>\epsilon\Bigg|X_n\in E_n(\eta/2)\right)\mathbb{P}(X_n\in E_n(\eta/2))\\
    &&+\mathbb{P}\left(\sup_{\stackrel{g_1,g_2\in\mathcal{G}}{\|g_1-g_2\|_{L_2}<\kappa}}|V_n(g_1)-V_n(g_2)|>\epsilon\Bigg|X_n\not\in E_n(\eta/2)\right)\mathbb{P}(X_n\not\in E_n(\eta/2))\\
    &\leq&\mathbb{P}\left(\sup_{\stackrel{g_1,g_2\in\mathcal{G}}{\|g_1-g_2\|_{L_2}<\kappa}}|V_n(g_1)-V_n(g_2)|>\epsilon\Bigg|X_n\in E_n(\eta/2)\right)+\mathbb{P}(X_n\not\in E_n(\eta/2))\\
    &\leq&\eta/2+\eta/2=\eta,
\end{eqnarray*}
which proves \cref{equicontinuity_V}.

\subsection{Convergence of finite-dimensional distributions}
The goal is to prove that
\[\sqrt{n}\int_{\Omega}(\hat{\delta}_n-\delta)(x)[g_1(x),\ldots,g_m(x)]dx\xrightarrow{\mathscr{L}} N\left(0,\left(\sigma^2\int_\Omega g_i(x)g_j(x)/p(x)\right)_{ij}\right).\]
Clearly, this is equivalent to proving that for any $\xi_1,\ldots,\xi_m\in\mathbb{R}$, 
\[\sqrt{n}\int_{\Omega}(\hat{\delta}_n-\delta)(x)\sum_{i=1}^m \xi_ig_i(x)dx\xrightarrow{\mathscr{L}} N\left(0,\sigma^2\int_\Omega\left(\sum_{i=1}^m\xi_ig_i(x)\right)^2\Big/p(x)dx\right).\]
By the Cram\'{e}r--Wold theorem, this implies that we only need to prove the one-dimensional result
\begin{equation}\label{CLT}
  \sqrt{n}\int_{\Omega}(\hat{\delta}_n-\delta)(x)g(x)dx\xrightarrow{\mathscr{L}} N\left(0,\sigma^2\int_\Omega g^2(x)/p(x)dx\right), 
\end{equation}
for each $g\in L_2$.

The limiting result \cref{CLT} has been partially established in the literature
\citep{mammen1997penalized,tuo2015efficient,wang2020semi}. However, the existing theorems do not cover the desired result for a general $g\in L_2$ and $\lambda\sim n^{-1}$, which is given in \Cref{Prop:CLT}.
It is worth noting that, when $g/p\in\mathcal{H}$, \Cref{Prop:CLT} requires a less stringent condition $n^{1/2}\lambda_n \rightarrow 0$, in contrast to $n\lambda_n=O(1)$ required by \Cref{Th:WeakConvergence}. According to a discussion made by \citet{tuo2024asymptotic}, this advantage is apparently limited to cases where $g/p\in\mathcal{H}$. In addition, the sub-Gaussian requirement for $e_i$'s is waived in \Cref{Prop:CLT}. 

\begin{theorem}\label{Prop:CLT}
Suppose Conditions \ref{C:truefunction}, \ref{C:input}, and \ref{C:tuning} are fulfilled, and the $e_i$'s are i.i.d., independent of the $x_i$, with $\mathbb{E}e_1=0$ and $\mathbb{E}e_1^2=\sigma^2\in(0,+\infty)$. Then \cref{CLT} holds for each $g\in L_2(\Omega)$. If $g/p\in\mathcal{H}$, the same conclusion holds when $n\lambda_n=O(1)$ in Condition \ref{C:tuning} is replaced by $n^{1/2}\lambda_n\rightarrow0$.
\end{theorem}

\begin{proof}
    Let $\tilde{g}:=g/p$ and $u_g:=\lambda^{-1}(v_g-\hat{v}_{g,X_n})$, so that $u_i(g)=n^{-1}u_g(x_i)$ in Property \ref{P2}. By Properties \ref{P1} and \ref{P2},
    \[
    |B_n(g)|\leq\sqrt{n}\lambda\|u_g\|_\mathcal{H}\|\delta\|_\mathcal{H},\qquad V_n(g)=\frac{1}{\sqrt{n}}\sum_{i=1}^n\tilde{g}(x_i)e_i+\frac{1}{\sqrt{n}}\sum_{i=1}^n(u_g-\tilde{g})(x_i)e_i .
    \]
    The first sum converges in distribution to $N(0,\sigma^2\int_\Omega g^2/p\,dx)$ by the central limit theorem, and, given $X_n$, the second sum has mean zero and variance $\sigma^2\|u_g-\tilde{g}\|_{X_n}^2$. Hence \cref{CLT} follows from
    \[
    \|u_g-\tilde{g}\|_{X_n}\xrightarrow{p}0\quad\text{and}\quad\sqrt{n}\lambda\|u_g\|_\mathcal{H}\xrightarrow{p}0,
    \]
    where the first implies, by Chebyshev's inequality, that the second sum tends to zero in probability, and the second implies $B_n(g)\xrightarrow{p}0$ by the bound on $|B_n(g)|$ above.

    We first prove these two limits when $\tilde{g}\in\mathcal{H}$, using only $n^{1/2}\lambda\rightarrow0$ and $n^{2m/d}\lambda\rightarrow\infty$; this also gives the last assertion of the theorem. Let $A_nv:=n^{-1}\sum_{i=1}^nK(\cdot,x_i)v(x_i)$ and $T_Pv:=\int_\Omega K(\cdot,x)v(x)p(x)dx=\mathbb{E}A_nv$. These are bounded, self-adjoint, and nonnegative operators on $\mathcal{H}$ with $\langle A_nv,v\rangle_\mathcal{H}=\|v\|_{X_n}^2$ and $\langle T_Pv,v\rangle_\mathcal{H}=\int_\Omega v^2p\,dx$, and $v_g=T_P\tilde{g}$. The first-order condition of $\hat{v}_{g,X_n}$ gives $u_g=(A_n+\lambda)^{-1}v_g$, so
    \[
    (A_n+\lambda)(\tilde{g}-u_g)=G+\lambda\tilde{g},\qquad G:=(A_n-T_P)\tilde{g}.
    \]
    This equation controls $u_g-\tilde{g}$ in the norm associated with $A_n+\lambda$,
    \[
    Q_n^2:=\langle(A_n+\lambda)(u_g-\tilde{g}),u_g-\tilde{g}\rangle_\mathcal{H}=\|u_g-\tilde{g}\|_{X_n}^2+\lambda\|u_g-\tilde{g}\|_\mathcal{H}^2,
    \]
    which bounds both quantities above: $\|u_g-\tilde{g}\|_{X_n}\leq Q_n$ and $\sqrt{n}\lambda\|u_g\|_\mathcal{H}\leq n^{1/2}\lambda\|\tilde{g}\|_\mathcal{H}+(n\lambda)^{1/2}Q_n$. Since $n^{1/2}\lambda\rightarrow0$, it suffices to show $Q_n\xrightarrow{p}0$ and $(n\lambda)^{1/2}Q_n\xrightarrow{p}0$. Both follow from $Q_n=O_\mathbb{P}(n^{-1/2}\lambda^{-a/2}+\lambda^{1/2})$ with $a:=d/(2m)<1$, which we now prove. By the equation,
    \[
    Q_n=\|(A_n+\lambda)^{-1/2}(G+\lambda\tilde{g})\|_\mathcal{H}\leq\|(A_n+\lambda)^{-1/2}G\|_\mathcal{H}+\lambda^{1/2}\|\tilde{g}\|_\mathcal{H}.
    \]
    By Theorem S4.1 of \citet{tuo2024asymptotic} and Conditions \ref{C:truefunction} and \ref{C:input}, for each $\eta>0$ there exist constants $C_1$ and $C_2$, depending on $\eta$, and events $\Xi_n$ with $\mathbb{P}(\Xi_n)\geq1-\eta$ on which $\|v\|_{L_2}\leq C_1\|v\|_{X_n}+C_2 n^{-m/d}\|v\|_\mathcal{H}$ for all $v\in\mathcal{H}$. As $n^{-2m/d}=o(\lambda)$, this gives $(A_n+\lambda)^{-1}\preceq C_3(T_P+\lambda)^{-1}$ on $\Xi_n$ for large $n$ and a constant $C_3$, so that $\|(A_n+\lambda)^{-1/2}G\|_\mathcal{H}^2=\langle(A_n+\lambda)^{-1}G,G\rangle_\mathcal{H}\leq C_3\|(T_P+\lambda)^{-1/2}G\|_\mathcal{H}^2$ on $\Xi_n$. To bound $\|(T_P+\lambda)^{-1/2}G\|_\mathcal{H}$, let $\langle u,v\rangle_\lambda:=\int_\Omega uvp\,dx+\lambda\langle u,v\rangle_\mathcal{H}=\langle(T_P+\lambda)u,v\rangle_\mathcal{H}$, under which $\mathcal{H}$ is a reproducing kernel Hilbert space with kernel $K_\lambda(\cdot,x):=(T_P+\lambda)^{-1}K(\cdot,x)$, and let $\zeta:=(T_P+\lambda)^{-1}G=n^{-1}\sum_{i=1}^n\{\tilde{g}(x_i)K_\lambda(\cdot,x_i)-\mathbb{E}[\tilde{g}(x_1)K_\lambda(\cdot,x_1)]\}$, an average of $n$ i.i.d.\ mean-zero random elements of $\mathcal{H}$. Substituting $u=(T_P+\lambda)^{1/2}v$, we obtain
    \begin{eqnarray*}
        \|(T_P+\lambda)^{-1/2}G\|_\mathcal{H}^2&=&\sup_{0\neq u\in\mathcal{H}}\frac{\langle(T_P+\lambda)^{-1/2}G,u\rangle_\mathcal{H}^2}{\|u\|_\mathcal{H}^2}\\
        &=&\sup_{0\neq v\in\mathcal{H}}\frac{\big[n^{-1}\sum_{i=1}^n\tilde{g}(x_i)v(x_i)-\mathbb{E}\{\tilde{g}(x_1)v(x_1)\}\big]^2}{\int_\Omega v^2p\,dx+\lambda\|v\|_\mathcal{H}^2}\\
        &=&\sup_{0\neq v\in\mathcal{H}}\frac{\langle\zeta,v\rangle_\lambda^2}{\|v\|_\lambda^2}=\|\zeta\|_\lambda^2,
    \end{eqnarray*}
    and, as the cross terms in $\mathbb{E}\|\zeta\|_\lambda^2$ vanish,
    \[
    \mathbb{E}\|\zeta\|_\lambda^2\leq\frac{1}{n}\mathbb{E}\|\tilde{g}(x_1)K_\lambda(\cdot,x_1)\|_\lambda^2=\frac{1}{n}\mathbb{E}\big[\tilde{g}^2(x_1)K_\lambda(x_1,x_1)\big]\leq\frac{1}{n}\int_\Omega\frac{g^2}{p}dx\,\sup_{x\in\Omega}K_\lambda(x,x).
    \]
    Finally, $K_\lambda(x,x)=\sup_{\|v\|_\lambda\leq1}v^2(x)$, and $\|v\|_\lambda\leq1$ implies $\|v\|_{L_2}\leq(\inf p)^{-1/2}$ and $\|v\|_\mathcal{H}\leq\lambda^{-1/2}$, so the interpolation inequality $\|v\|_{L_\infty}\leq A\|v\|_{L_2}^{1-a}\|v\|_{H^m}^{a}$ \citep[Theorem 1]{brezis2019sobolev} and Condition \ref{C:truefunction} give $\sup_{x\in\Omega}K_\lambda(x,x)\leq C_4\lambda^{-a}$. Hence $\mathbb{E}\|\zeta\|_\lambda^2\leq C_4n^{-1}\lambda^{-a}\int_\Omega g^2/p\,dx$, and Markov's inequality gives $\|\zeta\|_\lambda=O_\mathbb{P}(n^{-1/2}\lambda^{-a/2})$. As $\eta$ is arbitrary, the three bounds on $Q_n$, on $\|(A_n+\lambda)^{-1/2}G\|_\mathcal{H}$, and on $\|\zeta\|_\lambda$ give $Q_n=O_\mathbb{P}(n^{-1/2}\lambda^{-a/2}+\lambda^{1/2})$. Hence $Q_n\xrightarrow{p}0$, because $\lambda\rightarrow0$ and $n\lambda^{a}=(n^{2m/d}\lambda)^{a}\rightarrow\infty$, and $(n\lambda)^{1/2}Q_n=O_\mathbb{P}(\lambda^{(1-a)/2}+n^{1/2}\lambda)\xrightarrow{p}0$, because $a<1$ and $n^{1/2}\lambda\rightarrow0$.

    For a general $g\in L_2(\Omega)$, $\|\tilde{g}\|_\mathcal{H}$ may be infinite, so we approximate $g$ by $p\psi$ with $\psi\in\mathcal{H}$: given $\epsilon,\eta>0$, take $\psi\in\mathcal{H}$ with $\|g-p\psi\|_{L_2}<\epsilon$ and let $g_\epsilon:=g-p\psi$. Here we have used the fact that $p\mathcal{H}:=\{ph:h\in\mathcal{H}\}$ is dense in $L_2(\Omega)$. To see this, take any $h\in L_2(\Omega)$. According to Condition \ref{C:input}, $\|h/p\|_{L_2}\leq \|h\|_{L_2}/\inf_{x\in\Omega}p(x)<\infty$. It is known that $\mathcal{H}$ is dense in $L_2(\Omega)$ \citep{wendland2004scattered}. So we can find a sequence $\{s_1,s_2,\cdots\}\subset\mathcal{H}$ such that $\lim_{i\rightarrow\infty}\|s_i-h/p\|_{L_2}= 0$. Now use Condition \ref{C:input} again to obtain that $\|ps_i-h\|_{L_2}\leq \sup_{x\in\Omega}p(x)\|s_i-h/p\|_{L_2}\rightarrow 0$ as $i\rightarrow\infty$, which proves that $p\mathcal{H}$ is dense in $L_2(\Omega)$.
    By the case just proved, $B_n(p\psi)\xrightarrow{p}0$ and $Z_n(p\psi)\xrightarrow{\mathscr{L}}N(0,\sigma^2\int_\Omega p\psi^2dx)$, and $\int_\Omega p\psi^2dx\rightarrow\int_\Omega g^2/p\,dx$ as $\epsilon\rightarrow0$. If $X_n\in E_n(\eta)$, the set in Property \ref{P3}, then Properties \ref{P1}--\ref{P3} give $|B_n(g_\epsilon)|\leq\sqrt{n}C_\eta\lambda^{1/2}\epsilon\|\delta\|_\mathcal{H}$ and $\operatorname{Var}(V_n(g_\epsilon)|X_n)=\sigma^2\lambda^{-2}\|\hat{v}_{g_\epsilon,X_n}-v_{g_\epsilon}\|_{X_n}^2\leq\sigma^2C_\eta^2\epsilon^2$, as in \Cref{Sec:AsyEqui}. Since $n\lambda_n=O(1)$, letting $\epsilon\rightarrow0$ for fixed $\eta$ and then $\eta\rightarrow0$ shows that $B_n(g)=B_n(p\psi)+B_n(g_\epsilon)\xrightarrow{p}0$ and that $\lim_{\epsilon\rightarrow0}\limsup_{n\rightarrow\infty}\mathbb{P}(|Z_n(g)-Z_n(p\psi)|>\tau)=0$ for every $\tau>0$. As $|\mathbb{E}e^{\mathrm{i}tZ_n(g)}-\mathbb{E}e^{\mathrm{i}tZ_n(p\psi)}|\leq|t|\tau+2\mathbb{P}(|Z_n(g)-Z_n(p\psi)|>\tau)$, letting $n\rightarrow\infty$, then $\epsilon\rightarrow0$, and then $\tau\rightarrow0$ shows that the characteristic function of $Z_n(g)$ converges to that of $N(0,\sigma^2\int_\Omega g^2/p\,dx)$, which proves \cref{CLT}.
\end{proof}

With the asymptotic equicontinuity established in \Cref{Sec:AsyEqui} and the convergence of the finite-dimensional distributions in \Cref{Prop:CLT}, we have completed the proof of \Cref{Th:WeakConvergence}.

We also record the consistency of KRR used in \Cref{sec:D}: under Conditions \ref{C:truefunction}--\ref{C:tuning},
\begin{equation}\label{KRRL2}
    \|\hat{\delta}_n-\delta\|_{L_2}=o_\mathbb{P}(1).
\end{equation}
To see this, write $\hat{\delta}_n-\delta=(\tilde{\delta}_n-\delta)+u_n$, where $\tilde{\delta}_n:=\operatorname*{argmin}_{v\in\mathcal{H}}\|v-\delta\|_{X_n}^2+\lambda_n\|v\|_\mathcal{H}^2$ is the noiseless KRR of $\delta$ and $u_n:=K(\cdot,X_n)(K(X_n,X_n)+n\lambda_nI)^{-1}(e_1,\ldots,e_n)^T$. Comparing the objective of $\tilde{\delta}_n$ at $\tilde{\delta}_n$ and at $\delta$ gives $\|\tilde{\delta}_n-\delta\|_{X_n}\leq\lambda_n^{1/2}\|\delta\|_\mathcal{H}$ and $\|\tilde{\delta}_n\|_\mathcal{H}\leq\|\delta\|_\mathcal{H}$. For the noise part, with $\mathbf{S}_\lambda:=K(X_n,X_n)(K(X_n,X_n)+n\lambda_nI)^{-1}$, $\mathbb{E}[\|u_n\|_{X_n}^2|X_n]=\sigma^2n^{-1}\operatorname{tr}(\mathbf{S}_\lambda^2)\leq\sigma^2n^{-1}\operatorname{tr}(\mathbf{S}_\lambda)$ and $\mathbb{E}[\|u_n\|_\mathcal{H}^2|X_n]\leq\sigma^2(n\lambda_n)^{-1}\operatorname{tr}(\mathbf{S}_\lambda)$. With $A_n$, $T_P$, $\Xi_n$, $C_3$, and $C_4$ as in the proof of \Cref{Prop:CLT}, $\operatorname{tr}(\mathbf{S}_\lambda)=\operatorname{tr}\{A_n(A_n+\lambda_n)^{-1}\}=n^{-1}\sum_{i=1}^n\langle(A_n+\lambda_n)^{-1}K(\cdot,x_i),K(\cdot,x_i)\rangle_\mathcal{H}\leq C_3C_4\lambda_n^{-d/(2m)}$ on $\Xi_n$ for large $n$, by $(A_n+\lambda_n)^{-1}\preceq C_3(T_P+\lambda_n)^{-1}$ and $\langle(T_P+\lambda_n)^{-1}K(\cdot,x),K(\cdot,x)\rangle_\mathcal{H}=K_{\lambda_n}(x,x)\leq C_4\lambda_n^{-d/(2m)}$, shown in the proof of \Cref{Prop:CLT}. Hence $\|\tilde{\delta}_n-\delta\|_{X_n}$, $\|u_n\|_{X_n}$, and $n^{-m/d}(\|\tilde{\delta}_n-\delta\|_\mathcal{H}+\|u_n\|_\mathcal{H})$ tend to zero in probability under Condition \ref{C:tuning}, and the norm equivalence on $\Xi_n$ gives \cref{KRRL2}. For quasi-uniform fixed designs, see also Corollary 2.9 of \citet{tuo2020improved}.

\begin{remark}\label{rem:fixed}
    Condition \ref{C:input} assumes i.i.d.\ inputs. The proofs in this appendix use the randomness of the design only through the norm inequalities between $\|\cdot\|_{L_2}$ and $\|\cdot\|_{X_n}$ (Property \ref{P3} and the events $\Xi_n$ in the proof of \Cref{Prop:CLT}) and through the limit theorems for sums over the $x_i$. Let the design be deterministic, with an empirical distribution that converges weakly to a distribution with a density $p$ as in Condition \ref{C:input}, and let $\nu>2$. Then the conclusions of \Cref{Th:WeakConvergence,Th:limit} hold if (i) the designs are quasi-uniform, that is, the ratio of the fill distance $\sup_{x\in\Omega}\min_i\|x-x_i\|$ to the separation radius $\min_{i\neq j}\|x_i-x_j\|/2$ is bounded, and (ii) $\sup\{|n^{-1}\sum_{i=1}^n\varphi(x_i)-\int_\Omega\varphi p\,dx|:\|\varphi\|_{C^2(\Omega)}\leq1\}=o(\lambda_n^{1/2})$. Under (i), the norm inequalities hold for every $n$ \citep[Section S4.2]{tuo2024asymptotic}; under (ii), $\|\zeta\|_\lambda=o(1)$ in the proof of \Cref{Prop:CLT} for $g/p\in\mathcal{H}$, because $\|\tilde{g}v\|_{C^2(\Omega)}\leq C\lambda_n^{-1/2}$ if $\|v\|_\lambda\leq1$ and $\nu>2$, and the Lindeberg--Feller theorem replaces the central limit theorem. In one dimension, the quantile design $x_i=F^{-1}\{(i-1/2)/n\}$ of a distribution whose density is bounded away from zero and infinity satisfies (i), and also (ii) if $n\lambda_n^{1/2}\rightarrow\infty$, for example if $\lambda_n\asymp n^{-1}$, because the error of the midpoint rule is $O(n^{-1})$; for a uniform density, this is the midpoint grid. Simulations with these and other designs are given in \Cref{sec:J-designs}.
\end{remark}

\begin{remark}\label{rem:interp}
    Suppose that $y^s\in\mathcal{H}$ and that $\hat{y}^s$ is the interpolant in $\mathcal{H}$ of $N$ runs at quasi-uniform inputs, as in \Cref{rem:surrogate}. The error $\hat{y}^s-y^s$ vanishes at the runs and has $\mathcal{H}$-norm at most $\|y^s\|_\mathcal{H}$, so the norm inequalities of \citet[Section S4]{tuo2024asymptotic} give $\|\hat{y}^s-y^s\|_{X_n}=O_\mathbb{P}(\max\{N^{-m/d},n^{-m/d}\})\|y^s\|_\mathcal{H}$ under Conditions \ref{C:truefunction} and \ref{C:input}. Hence the tests with $\hat{y}^s$ and with $y^s$ are asymptotically equivalent if $N/n^{d/(2m)}\rightarrow\infty$; for $d=1$ and $\nu=3.5$, $N$ need only grow faster than $n^{1/8}$. This does not cover extrapolation, which \Cref{sec:K} checks numerically for the shear-layer data.
\end{remark}

\section{The calibrated test}\label{sec:C}

This appendix gives the covariance matrix $\Sigma_\Delta$ used in \Cref{sec:calibration}, the full algorithm, and the asymptotic theory of the calibrated test. Write $\dot{y}:=\frac{\partial y^s}{\partial\theta}(\cdot,\theta^*)$ for the column vector of $\theta$-derivatives and $\ddot{y}:=\frac{\partial^2 y^s}{\partial\theta\,\partial\theta^T}(\cdot,\theta^*)$. Fix a test domain $\Delta$ with orthonormal basis $\{h_j\}_{j\geq1}$, extended by zero outside $\Delta$. To first order, replacing $\theta^*$ by $\hat{\theta}$ adds $-a_j^T\sqrt{n}(\hat{\theta}-\theta^*)$ to $\sqrt{n}\hat{s}_j$, where $a_j:=\int_\Delta\dot{y}\sqrt{p}\,h_j\,dx$. Let $\tilde{a}_j:=\int_\Delta\dot{y}\,p^{-1/2}h_j\,dx$, and let
\begin{equation}\label{eq:M}
    M_\theta:=\int_\Omega\dot{y}\dot{y}^T dx-\int_\Omega\delta\,\ddot{y}\,dx,\qquad
    \tilde{\Gamma}:=\int_\Omega\dot{y}\dot{y}^T p^{-1}dx .
\end{equation}
Under $H_0$ the second integral in $M_\theta$ vanishes, and $M_\theta$ is the Gram matrix of the components of $\dot{y}$ in $L_2(\Omega)$. \Cref{Th:cal} below shows that, if $f=y^s(\cdot,\theta^*)$ on $\Delta$, the vector $(\sqrt{n}\rho_j\hat{s}_j)_{j\geq1}$ converges weakly in $l_\infty$ to a normal distribution with mean zero and covariance $\sigma^2\Sigma_\Delta$, whose $(j,j')$ entry is
\begin{equation}\label{eq:Sigma}
    (\Sigma_\Delta)_{jj'}:=\rho_j\rho_{j'}\big(\mathbb{I}\{j=j'\}-a_j^T M_\theta^{-1}\tilde{a}_{j'}-\tilde{a}_j^T M_\theta^{-1}a_{j'}+a_j^T M_\theta^{-1}\tilde{\Gamma}M_\theta^{-1}a_{j'}\big),\qquad j,j'\geq1 .
\end{equation}
The two middle terms come from the correlation between $\hat{s}_j$ and $\hat{\theta}$, and the last from the variance of $\hat{\theta}$. The estimate $\hat{\Sigma}_\Delta$ is obtained by evaluating $a_j$, $\tilde{a}_j$, $M_\theta$, and $\tilde{\Gamma}$ at $\hat{\theta}$ with $p$ and $f$ replaced by $\hat{p}_n$ and $\hat{f}_n$. Since $\rho_j\rightarrow0$ and the matrix of the bracketed terms is the identity plus a matrix of rank at most $2d_\theta$, $\Sigma_\Delta$ is a compact operator on $l_2$, and it is positive semidefinite by the proof of \Cref{Prop:cal-joint}. The same holds for $\hat{\Sigma}_\Delta$ when $\hat{p}_n>0$ and the estimate of $M_\theta$ is invertible. Both therefore admit orthonormal bases of eigenvectors with nonnegative eigenvalues.

We do not divide the coordinates in \cref{eq:cal-T} by $\sqrt{\hat{\tau}_k}$. For the global test, $\Sigma_\Omega$ has eigenvalues close to zero in the directions of the coefficients of $\dot{y}p^{-1/2}$; along these directions the coefficients carry little noise and, by \Cref{Lem:null}, no signal. Their factors in \cref{eq:cal-T} are then close to one, whereas standardizing would amplify the estimation error. \Cref{alg:cal-fmmt} states the procedure in full.

\begin{algorithm}[!htbp]
\caption{Calibrated Fourier maximum modulus test}
\label{alg:cal-fmmt}
\small
\begin{algorithmic}[1]
\REQUIRE Data $\{(x_l,y_l)\}_{l=1}^n$; simulator $y^s(\cdot,\theta)$ with its first two $\theta$-derivatives; test domains $\Delta\in\{\Omega,\Omega_1,\ldots,\Omega_N\}$, each with basis functions $h_1,\ldots,h_{K_\Delta}$ and weights $\rho_1,\ldots,\rho_{K_\Delta}$; level $\alpha$.
\STATE Fit $\hat{f}_n$, $\hat{\sigma}_n$, and $\hat{p}_n$ as in \Cref{sec:details}.
\STATE Compute $\hat{\theta}$ by \cref{eq:thetastar}.
\FOR{each test domain $\Delta$}
\STATE Compute $\hat{s}_j$ by \cref{eq:shat-cal} and $\hat{\Sigma}_\Delta$, and the eigenpairs $(\hat{\tau}_k,\hat{u}_k)$ of $\hat{\Sigma}_\Delta$.
\STATE Compute $T_\Delta$ and $p_\Delta$ by \cref{eq:cal-T}.
\ENDFOR
\STATE Reject $H_0$ if $p_\Omega\leq\alpha$, and reject $H_{0i}$ if $p_{\Omega_i}\leq\alpha/N$.
\end{algorithmic}
\end{algorithm}

For a test domain $\Delta$, write $H_{0\Delta}$ for $H_0$ if $\Delta=\Omega$ and for $H_{0i}$ if $\Delta=\Omega_i$, and define $H_{1\Delta}$ likewise.

\begin{theorem}\label{Th:cal}
    Suppose Conditions \ref{C:truefunction}--\ref{C:ident} hold and $\|\sqrt{\hat{p}_n}-\sqrt{p}\|_{L_\infty}\xrightarrow{p}0$. Let $\Delta$ be a test domain.
    \begin{enumerate}[label=\rm{(\roman*)}]
        \item Under $H_{0\Delta}$, the infinite-dimensional vector $(\sqrt{n}\rho_1\hat{s}_1,\sqrt{n}\rho_2\hat{s}_2,\ldots)^T$, as a random element taking values in $l_\infty$, converges weakly to a normal distribution with mean zero and covariance matrix $\sigma^2\Sigma_\Delta$. For $\Delta=\Omega_i$ this holds whether or not $H_0$ holds.
        \item Under $H_{1\Delta}$, $p_\Delta\xrightarrow{p}0$.
        \item Let $K_\Delta$ be fixed, let $\hat{\Sigma}_{\Delta,K}$ and $\Sigma_{\Delta,K}$ be the leading $K_\Delta\times K_\Delta$ blocks of $\hat{\Sigma}_\Delta$ and $\Sigma_\Delta$, and let $p_{\Delta,K}$ be computed by \cref{eq:cal-T} from $\hat{s}_1,\ldots,\hat{s}_{K_\Delta}$ and $\hat{\Sigma}_{\Delta,K}$, as in \Cref{alg:cal-fmmt}. If $\Sigma_{\Delta,K}\neq0$ has simple positive eigenvalues, then under $H_{0\Delta}$, $\lim_{n\rightarrow\infty}\mathbb{P}(p_{\Delta,K}\leq\alpha)=\alpha$ for each $\alpha\in(0,1)$.
    \end{enumerate}
\end{theorem}

In addition to Conditions \ref{C:truefunction}--\ref{C:decay}, \Cref{Th:cal} requires the following conditions.
\begin{enumerate}[label=\rm{C}\arabic*,   
                  ref=\rm{C}\arabic*,     
                  start=7]
    \item (Simulator smoothness) The map $y^s$ is continuous on $\Omega\times\Theta$; for each $x\in\Omega$, $y^s(x,\cdot)\in C^{2,1}(\Theta)$, and $\frac{\partial y^s}{\partial\theta}$ and $\frac{\partial^2 y^s}{\partial\theta\,\partial\theta^T}$ are uniformly bounded on $\Omega\times\Theta$. On a neighborhood of $\theta^*$, the map $\theta\mapsto y^s(\cdot,\theta)$ takes values in $\mathcal{H}$ and is twice continuously differentiable as a map into $\mathcal{H}$.\label{C:simsmooth}
    \item (Identifiability) The minimizer $\theta^*$ in \cref{eq:thetastar} exists, is unique, and lies in the interior of $\Theta$, and the matrix $M_\theta$ in \cref{eq:M} is positive definite.\label{C:ident}
\end{enumerate}
The proof is given in \Cref{sec:D}.

\section{Proof of \texorpdfstring{\Cref{Th:cal}}{Theorem C.1}}\label{sec:D}

The notation is that of \Cref{sec:calibration,sec:C}, and $Z_n$ is the process of \Cref{sec:A} with $y^s\equiv0$, so that $Z_n(g)=\sqrt{n}\int_\Omega(\hat{f}_n-f)(x)g(x)dx$; likewise, \cref{KRRL2} is used with $y^s\equiv0$. Write $\dot{y}_\vartheta:=\frac{\partial y^s}{\partial\theta}(\cdot,\vartheta)$, so that $\dot{y}=\dot{y}_{\theta^*}$. By Conditions \ref{C:truefunction} and \ref{C:simsmooth}, $f=y^s(\cdot,\theta^*)+\delta\in\mathcal{H}$. Let $B_0:=\sup_{\Omega\times\Theta}\|\frac{\partial y^s}{\partial\theta}\|$ and $B_2:=\sup_{\Omega\times\Theta}\|\frac{\partial^2 y^s}{\partial\theta\,\partial\theta^T}\|$, which are finite by Condition \ref{C:simsmooth}, and let $\Theta_0\subset\Theta$ be a closed ball centered at $\theta^*$. Under the hypotheses of \Cref{Th:cal}, $\sqrt{\hat{p}_n}$ lies in any $L_\infty$-neighborhood of $\sqrt{p}$ with probability tending to one. We may therefore intersect the class $\mathcal{Q}$ of Condition \ref{C:density} with such a neighborhood and replace it by its closure in $L_\infty$, which contains $\sqrt{p}$, has the same entropy integral, and, by Condition \ref{C:input}, satisfies $q^2\geq\underline{p}:=\inf_{x\in\Omega}p(x)/2>0$ on $\Omega$ for every $q\in\mathcal{Q}$.

Since $\theta^*$ is an interior minimizer of $\int_\Omega(f-y^s(\cdot,\theta))^2dx$ by Condition \ref{C:ident}, and $f$, $y^s$, and $\frac{\partial y^s}{\partial\theta}$ are bounded, differentiating under the integral sign gives
\begin{equation}\label{eq:orth}
    \int_\Omega \delta\,\dot{y}\,dx=\mathbf{0}.
\end{equation}

\subsection{The calibration estimator}\label{sec:D-theta}

\begin{lemma}\label{Lem:theta-lin}
    Under Conditions \ref{C:truefunction}--\ref{C:tuning}, \ref{C:simsmooth}, and \ref{C:ident}, whether or not $H_0$ holds, $\hat{\theta}\xrightarrow{p}\theta^*$ and
    \[
    \sqrt{n}(\hat{\theta}-\theta^*)=M_\theta^{-1}Z_n(\dot{y})+o_\mathbb{P}(1)\xrightarrow{\mathscr{L}}N\big(0,\sigma^2M_\theta^{-1}\tilde{\Gamma}M_\theta^{-1}\big).
    \]
\end{lemma}

\begin{proof}
    Consistency follows as in the proof of Theorem 1 of \citet{tuo2015efficient}: since $\|\hat{f}_n-f\|_{L_2}\xrightarrow{p}0$ by \cref{KRRL2} and $\sup_\theta\|f-y^s(\cdot,\theta)\|_{L_2}<\infty$, the criterion defining $\hat{\theta}$ in \cref{eq:thetastar} converges to that defining $\theta^*$ uniformly on the compact set $\Theta$, and $\theta^*$ is the unique minimizer of the limit.

    With probability tending to one $\hat{\theta}$ is an interior point of $\Theta_0$, and the first-order condition for $\hat{\theta}$ reads $\int_\Omega(\hat{f}_n-y^s(\cdot,\hat{\theta}))\dot{y}_{\hat{\theta}}\,dx=\mathbf{0}$. Writing $\hat{f}_n-y^s(\cdot,\hat{\theta})=(\hat{f}_n-f)+\delta+(y^s(\cdot,\theta^*)-y^s(\cdot,\hat{\theta}))$, we obtain
    \[
    \mathbf{0}=\frac{1}{\sqrt{n}}Z_n(\dot{y}_{\hat{\theta}})+\int_\Omega \delta\,\dot{y}_{\hat{\theta}}\,dx+\int_\Omega\big(y^s(\cdot,\theta^*)-y^s(\cdot,\hat{\theta})\big)\dot{y}_{\hat{\theta}}\,dx .
    \]
    By Taylor's theorem and Condition \ref{C:simsmooth}, the second term is $\int_\Omega \delta\,\dot{y}\,dx+\int_\Omega \delta\,\ddot{y}\,dx\,(\hat{\theta}-\theta^*)+O(\|\hat{\theta}-\theta^*\|^2)$, and the third is $-\int_\Omega \dot{y}\dot{y}^T dx\,(\hat{\theta}-\theta^*)+O(\|\hat{\theta}-\theta^*\|^2)$. By \cref{eq:orth}, their sum is $-M_\theta(\hat{\theta}-\theta^*)+O(\|\hat{\theta}-\theta^*\|^2)$, and hence
    \begin{equation}\label{eq:theta-fo}
    \big(M_\theta+O_\mathbb{P}(\|\hat{\theta}-\theta^*\|)\big)\sqrt{n}(\hat{\theta}-\theta^*)=Z_n(\dot{y}_{\hat{\theta}}).
    \end{equation}
    The class $\{\frac{\partial y^s}{\partial\theta_k}(\cdot,\vartheta):\vartheta\in\Theta_0,\,k\leq d_\theta\}$ is bounded in $L_2$ by $|\Omega|^{1/2}B_0$ and is $|\Omega|^{1/2}B_2$-Lipschitz in $\vartheta$, so its $L_2$ covering numbers are at most $d_\theta(C_1/\epsilon)^{d_\theta}$ for some $C_1$ and all $\epsilon<C_1$, and it satisfies \cref{entropyintegral}. By \Cref{Th:WeakConvergence}, $Z_n$ is asymptotically uniformly $L_2$-equicontinuous in probability on this class \citep[Theorem 1.5.7]{VaartWCofEP00}. Since $\|\dot{y}_{\hat{\theta}}-\dot{y}\|_{L_2}\leq|\Omega|^{1/2}B_2\|\hat{\theta}-\theta^*\|\xrightarrow{p}0$, we have $Z_n(\dot{y}_{\hat{\theta}})=Z_n(\dot{y})+o_\mathbb{P}(1)=O_\mathbb{P}(1)$. As $M_\theta$ is invertible, \cref{eq:theta-fo} gives $\sqrt{n}(\hat{\theta}-\theta^*)=O_\mathbb{P}(1)$ and then the stated expansion. Finally, $Z_n(\dot{y})\xrightarrow{\mathscr{L}}N(0,\sigma^2\tilde{\Gamma})$ by \Cref{Th:WeakConvergence}.
\end{proof}

\Cref{Lem:theta-lin} extends Theorem 1 of \citet{tuo2015efficient}, who take the inputs uniform on $\Omega$, to a general input density. For uniform inputs, their covariance $4\sigma^2V^{-1}WV^{-1}$, with $V:=\mathbb{E}\big[\frac{\partial^2}{\partial\theta\,\partial\theta^T}(f-y^s)^2(x_1,\theta^*)\big]=2M_\theta/|\Omega|$ and $W:=\mathbb{E}[\dot{y}\dot{y}^T(x_1)]=\tilde{\Gamma}/|\Omega|^2$, equals $\sigma^2M_\theta^{-1}\tilde{\Gamma}M_\theta^{-1}$. For a general $p$, the factor $p^{-1}$ in $\tilde{\Gamma}$ comes from the first-order condition of KRR, as in the proof of \Cref{Prop:CLT}.

\subsection{The joint limit of the coefficients}\label{sec:D-joint}

\begin{proposition}\label{Prop:cal-joint}
    Suppose Conditions \ref{C:truefunction}--\ref{C:ident} hold, $\|\sqrt{\hat{p}_n}-\sqrt{p}\|_{L_\infty}\xrightarrow{p}0$, and $f=y^s(\cdot,\theta^*)$ on $\Delta$. Then, uniformly in $j\geq1$,
    \[
    \sqrt{n}\rho_j\hat{s}_j=Z_n(\varphi_j)+o_\mathbb{P}(1),\qquad \varphi_j:=\rho_j\big(h_j\sqrt{p}-a_j^T M_\theta^{-1}\dot{y}\big),
    \]
    and $(\sqrt{n}\rho_j\hat{s}_j)_{j\geq1}$ converges weakly in $l_\infty$ to $N(0,\sigma^2\Sigma_\Delta)$ with $\Sigma_\Delta$ given by \cref{eq:Sigma}.
\end{proposition}

\begin{proof}
    Let $\chi_j:=\rho_jh_j\sqrt{p}$ and $\hat{\chi}_j:=\rho_jh_j\sqrt{\hat{p}_n}$, with $h_j$ extended by zero outside $\Delta$. We work on the event $\{\hat{\theta}\in\Theta_0,\ \sqrt{\hat{p}_n}\in\mathcal{Q}\}$, whose probability tends to one. Let $\mathcal{S}_nv:=K(\cdot,X_n)(K(X_n,X_n)+n\lambda_nI)^{-1}v(X_n)$ be the noiseless KRR of a function $v$, so that $\hat{\delta}_n^{\hat{\theta}}=\hat{f}_n-\mathcal{S}_ny^s(\cdot,\hat{\theta})$. Since $f=y^s(\cdot,\theta^*)$ on $\Delta$, a Taylor expansion of $y^s(\cdot,\hat{\theta})$ about $\theta^*$ gives
    \[
    \sqrt{n}\rho_j\hat{s}_j=Z_n(\hat{\chi}_j)-\rho_j\bar{a}_j^T\sqrt{n}(\hat{\theta}-\theta^*)+\eta_j,\qquad \bar{a}_j:=\int_\Delta \dot{y}\sqrt{\hat{p}_n}h_j\,dx,
    \]
    where $\eta_j$ consists of the second-order Taylor term and $\sqrt{n}\rho_j\int_\Delta(I-\mathcal{S}_n)y^s(\cdot,\hat{\theta})\sqrt{\hat{p}_n}h_j\,dx$. The Taylor term is at most $\frac{\sqrt{n}}{2}\sup_j\rho_jB_2\|\hat{\theta}-\theta^*\|^2\|\sqrt{\hat{p}_n}\|_{L_2}=O_\mathbb{P}(n^{-1/2})$ by \Cref{Lem:theta-lin}. By Condition \ref{C:simsmooth}, $(I-\mathcal{S}_n)y^s(\cdot,\hat{\theta})=(I-\mathcal{S}_n)\{y^s(\cdot,\theta^*)+\dot{y}^T(\hat{\theta}-\theta^*)+r\}$ with $\|r\|_\mathcal{H}=O_\mathbb{P}(\|\hat{\theta}-\theta^*\|^2)$. The first part gives $-B_n(\rho_jh_j\sqrt{\hat{p}_n})$, where $B_n$ is the bias term of \Cref{Sec:AsyEqui} for the regression function $y^s(\cdot,\theta^*)$; it tends to zero uniformly in $j$ by \cref{equicontinuity_B1} and $B_n(g)\xrightarrow{p}0$ for each $g\in L_2(\Omega)$, shown in the proof of \Cref{Prop:CLT}. The second part is at most $\sup_j\rho_j\|\sqrt{\hat{p}_n}\|_{L_\infty}\sqrt{n}\|\hat{\theta}-\theta^*\|\|(I-\mathcal{S}_n)\dot{y}\|_{L_2}=o_\mathbb{P}(1)$, since $\|(I-\mathcal{S}_n)\dot{y}\|_{L_2}\xrightarrow{p}0$ by the argument for \cref{KRRL2}. The third part is $O_\mathbb{P}(\sqrt{n}\|r\|_\mathcal{H})=O_\mathbb{P}(n^{-1/2})$, since $\|\mathcal{S}_nr\|_\mathcal{H}\leq\|r\|_\mathcal{H}$ and $\mathcal{H}$ is continuously embedded in $L_\infty(\Omega)$. Hence $\sup_j|\eta_j|=o_\mathbb{P}(1)$. Moreover, $\sup_j\|\bar{a}_j-a_j\|\leq\sqrt{d_\theta}B_0\|\sqrt{\hat{p}_n}-\sqrt{p}\|_{L_2}\xrightarrow{p}0$. The class $\{\rho_jh_jq:j\geq1,\ q\in\mathcal{Q}\}$ satisfies \cref{entropyintegral} as in the proof of \Cref{Th:limit}, and $\sup_j\|\hat{\chi}_j-\chi_j\|_{L_2}\leq\sup_j\rho_j\|\sqrt{\hat{p}_n}-\sqrt{p}\|_{L_\infty}\xrightarrow{p}0$, so $\sup_j|Z_n(\hat{\chi}_j)-Z_n(\chi_j)|=o_\mathbb{P}(1)$ by asymptotic equicontinuity. Combining these with \Cref{Lem:theta-lin} and the linearity of $Z_n$, uniformly in $j$,
    \[
    \sqrt{n}\rho_j\hat{s}_j=Z_n(\chi_j)-\rho_ja_j^T M_\theta^{-1}Z_n(\dot{y})+o_\mathbb{P}(1)=Z_n(\varphi_j)+o_\mathbb{P}(1).
    \]
    The class $\mathcal{G}':=\{\chi_j:j\geq1\}\cup\{\frac{\partial y^s}{\partial\theta_k}(\cdot,\theta^*):k\leq d_\theta\}$ also satisfies \cref{entropyintegral}, so $Z_n$ converges weakly in $L_\infty(\mathcal{G}')$ by \Cref{Th:WeakConvergence}. Since $\sup_j\rho_j\|a_j\|\leq\sup_j\rho_j\|\dot{y}\sqrt{p}\|_{L_2}<\infty$, the map $z\mapsto(z(\chi_j)-\rho_ja_j^T M_\theta^{-1}z(\dot{y}))_{j\geq1}$ from $L_\infty(\mathcal{G}')$ to $l_\infty$ is linear and continuous. By the continuous mapping theorem, $(Z_n(\varphi_j))_{j\geq1}$, and hence $(\sqrt{n}\rho_j\hat{s}_j)_{j\geq1}$, converges weakly in $l_\infty$ to a centered normal distribution with covariance $\sigma^2\int_\Omega\varphi_j\varphi_{j'}/p\,dx$. Since $\{h_j\}$ is orthonormal in $L_2(\Delta)$, $\int_\Omega \chi_j\chi_{j'}/p\,dx=\rho_j\rho_{j'}\mathbb{I}\{j=j'\}$, $\int_\Omega \chi_j\dot{y}^T/p\,dx=\rho_j\tilde{a}_j^T$, and $\int_\Omega \dot{y}\dot{y}^T/p\,dx=\tilde{\Gamma}$. Hence
    \[
    \int_\Omega\frac{\varphi_j\varphi_{j'}}{p}dx=\rho_j\rho_{j'}\big[\mathbb{I}\{j=j'\}-a_j^T M_\theta^{-1}\tilde{a}_{j'}-\tilde{a}_j^T M_\theta^{-1}a_{j'}+a_j^T M_\theta^{-1}\tilde{\Gamma}M_\theta^{-1}a_{j'}\big],
    \]
    which is the $(j,j')$ entry of $\Sigma_\Delta$.
\end{proof}

\begin{lemma}\label{Lem:null}
    Under Conditions \ref{C:input}, \ref{C:simsmooth}, and \ref{C:ident}, for every test domain $\Delta$, the vector $(\rho_js_j)_{j\geq1}$ with $s_j:=\int_\Delta \delta\sqrt{p}\,h_j\,dx$ lies in $l_2$ and is orthogonal to the null space of $\Sigma_\Delta$.
\end{lemma}

\begin{proof}
    By Bessel's inequality, $\sum_js_j^2\leq\|\delta\sqrt{p}\|_{L_2}^2$ and $\sum_j\|a_j\|^2\leq\|\dot{y}\sqrt{p}\|_{L_2}^2$, so for $v\in l_2$ the series $\sum_jv_j\varphi_j$ converges in $L_2(\Omega)$. By the proof of \Cref{Prop:cal-joint}, $v^T\Sigma_\Delta v=\int_\Omega(\sum_jv_j\varphi_j)^2/p\,dx$ for $v\in l_2$. Since $\Sigma_\Delta$ is positive semidefinite and $p$ is bounded, $\Sigma_\Delta v=0$ if and only if $\sum_jv_j\varphi_j=0$ almost everywhere, that is, $\sum_jv_j\rho_jh_j\sqrt{p}=b^T M_\theta^{-1}\dot{y}$ with $b:=\sum_jv_j\rho_ja_j$. For such $v$, by \cref{eq:orth},
    \[
    \sum_jv_j\rho_js_j=\int_\Omega \delta\sum_jv_j\rho_jh_j\sqrt{p}\,dx=b^T M_\theta^{-1}\int_\Omega \delta\,\dot{y}\,dx=0 .
    \]
\end{proof}

\begin{lemma}\label{Lem:psi-hat}
    Under the hypotheses of \Cref{Th:cal}, whether or not $H_{0\Delta}$ holds, $\hat{\Sigma}_\Delta\xrightarrow{p}\Sigma_\Delta$ in the operator norm on $l_2$.
\end{lemma}

\begin{proof}
    Let $\hat{a}_j$, $\hat{\tilde{a}}_j$, $\hat{M}_\theta$, and $\hat{\Gamma}$ denote the estimates of $a_j$, $\tilde{a}_j$, $M_\theta$, and $\tilde{\Gamma}$ that define $\hat{\Sigma}_\Delta$. By Condition \ref{C:simsmooth}, Bessel's inequality, and the Cauchy--Schwarz inequality, $(\sum_j\|\hat{a}_j-a_j\|^2)^{1/2}$, $(\sum_j\|\hat{\tilde{a}}_j-\tilde{a}_j\|^2)^{1/2}$, and each entry of $\hat{M}_\theta-M_\theta$ and $\hat{\Gamma}-\tilde{\Gamma}$ are bounded by a constant multiple of $\|\hat{\theta}-\theta^*\|+\|\sqrt{\hat{p}_n}-\sqrt{p}\|_{L_\infty}+\|\hat{f}_n-f\|_{L_2}$; for the entries involving $\hat{p}_n^{-1/2}$ and $\hat{p}_n^{-1}$ we also use the lower bound $q^2\geq\underline{p}$ on $\mathcal{Q}$. This bound tends to zero in probability by \Cref{Lem:theta-lin}, the hypothesis on $\hat{p}_n$, and \cref{KRRL2}. Since $\sup_j\rho_j<\infty$, the map from $((a_j)_{j\geq1},(\tilde{a}_j)_{j\geq1},M_\theta,\tilde{\Gamma})$ to $\Sigma_\Delta$ in \cref{eq:Sigma} is continuous in the operator norm where $M_\theta$ is invertible.
\end{proof}

\subsection{Proof of \texorpdfstring{\Cref{Th:cal}}{Theorem C.1}}\label{sec:D-proof}

Fix $\Delta$ and write $\zeta_n:=(\sqrt{n}\rho_j\hat{s}_j)_{j\geq1}$, which lies in $l_2$ by Bessel's inequality, so that $T_\Delta=\sup_k|\langle\hat{u}_k,\zeta_n\rangle|/\hat{\sigma}_n$, where $\langle\cdot,\cdot\rangle$ and $\|\cdot\|$ denote the inner product and norm of $l_2$. As in \Cref{Prop:cal-joint}, we work on the event $\{\hat{\theta}\in\Theta_0,\ \sqrt{\hat{p}_n}\in\mathcal{Q}\}$, whose probability tends to one.

\begin{proof}[Proof of \Cref{Th:cal}]
    (i) Under $H_{0\Delta}$, $f=y^s(\cdot,\theta^*)$ on $\Delta$; for $\Delta=\Omega$ this uses the uniqueness of $\theta^*$. The claim is then \Cref{Prop:cal-joint}, which, like \Cref{Lem:theta-lin}, does not require $H_0$.

    (ii) Let $w_n:=(\rho_j\int_\Delta \delta\sqrt{\hat{p}_n}h_j\,dx)_{j\geq1}$ and $w:=(\rho_js_j)_{j\geq1}$, with $s_j$ as in \Cref{Lem:null}. By Bessel's inequality, $\|w_n-w\|\leq\sup_j\rho_j\|\delta\|_{L_2}\|\sqrt{\hat{p}_n}-\sqrt{p}\|_{L_\infty}\xrightarrow{p}0$. Under $H_{1\Delta}$, $\delta$ is continuous by Conditions \ref{C:truefunction} and \ref{C:simsmooth} and is not identically zero on $\Delta$, so $\delta\sqrt{p}\neq0$ in $L_2(\Delta)$ and, since $\{h_j\}$ is complete, $w\neq0$. For $v\in l_2$, let $g_v:=\sum_jv_j\rho_jh_j$, so that $\|g_v\|_{L_2}\leq\sup_j\rho_j\|v\|$. Writing $\hat{\delta}_n^{\hat{\theta}}=(\hat{f}_n-f)+\delta+(y^s(\cdot,\theta^*)-y^s(\cdot,\hat{\theta}))+(I-\mathcal{S}_n)y^s(\cdot,\hat{\theta})$, we obtain
    \begin{align*}
    \sup_{\|v\|\leq1}\big|\langle v,\zeta_n\rangle-\sqrt{n}\langle v,w_n\rangle\big|&\leq\sqrt{n}\sup_j\rho_j\|\sqrt{\hat{p}_n}\|_{L_\infty}\\
    &\quad\times\big(\|\hat{f}_n-f\|_{L_2}+|\Omega|^{1/2}B_0\|\hat{\theta}-\theta^*\|+\|(I-\mathcal{S}_n)y^s(\cdot,\hat{\theta})\|_{L_2}\big)\\
    &=o_\mathbb{P}(\sqrt{n})
    \end{align*}
    by \cref{KRRL2}, \Cref{Lem:theta-lin}, and the proof of \Cref{Prop:cal-joint}.

    By \Cref{Lem:null}, $w$ is orthogonal to the null space of the compact self-adjoint operator $\Sigma_\Delta$. Hence $w$ has a nonzero projection $\Pi w$ onto the eigenspace of some eigenvalue $\tau>0$ of $\Sigma_\Delta$; this eigenspace has finite dimension $r$. Choose $\gamma>0$ such that $\Sigma_\Delta$ has no other eigenvalue in $[\tau-2\gamma,\tau+2\gamma]$. By \Cref{Lem:psi-hat} and Weyl's inequality, with probability tending to one, $\hat{\Sigma}_\Delta$ has exactly $r$ eigenvalues in $(\tau-\gamma,\tau+\gamma)$, counted with multiplicity, and no other eigenvalue within distance $\gamma$ of $\tau$. Let $\hat{\Pi}$ be the orthogonal projection onto the span of the corresponding eigenvectors $\hat{u}_k$. The $\sin\theta$ theorem of \citet{davis1970rotation} gives $\|\hat{\Pi}-\Pi\|\leq\|\hat{\Sigma}_\Delta-\Sigma_\Delta\|/\gamma\xrightarrow{p}0$, and hence $\|\hat{\Pi}w_n\|\xrightarrow{p}\|\Pi w\|>0$. Therefore
    \[
    \sqrt{r}\,\hat{\sigma}_nT_\Delta\geq\|\hat{\Pi}\zeta_n\|=\sup_{v\in\hat{\Pi}l_2,\,\|v\|=1}\langle v,\zeta_n\rangle\geq\sqrt{n}\|\hat{\Pi}w_n\|-o_\mathbb{P}(\sqrt{n}),
    \]
    and $T_\Delta\xrightarrow{p}\infty$ by the consistency of $\hat{\sigma}_n$.

    Given the data, $W^2\sim\kappa_n\chi^2_{r_n}/r_n$. The eigenvalues of $S$ lie in $[0,1)$, so $\operatorname{tr}(D)\geq n-2\operatorname{tr}(S)$ and $\operatorname{tr}(D^2)\leq\operatorname{tr}(D)\leq n$, and $\operatorname{tr}(S)=O_\mathbb{P}(\lambda_n^{-d/(2m)})=o_\mathbb{P}(n)$ by the bound on $\operatorname{tr}(\mathbf{S}_\lambda)$ after \cref{KRRL2} and Condition \ref{C:tuning}. Hence $r_n\rightarrow\infty$ and $\kappa_n\rightarrow1$ in probability, and $W$ concentrates at one.

    It remains to bound the factors of $p_\Delta$. As in the proof of \Cref{Prop:cal-joint}, $v^T\hat{\Sigma}_\Delta v=\int_\Omega(\sum_jv_j\hat{\varphi}_j)^2/\hat{p}_n\,dx$ for $v\in l_2$, where $\hat{\varphi}_j:=\rho_j(h_j\sqrt{\hat{p}_n}-\hat{a}_j^T\hat{M}_\theta^{-1}\dot{y}_{\hat{\theta}})$ and $\hat{a}_j$ and $\hat{M}_\theta$ are as in the proof of \Cref{Lem:psi-hat}. Hence $v^T\hat{\Sigma}_\Delta v\leq \bar{C}_n\sum_j\rho_j^2v_j^2$ with $\bar{C}_n:=2+2\|\hat{M}_\theta^{-1}\hat{\Gamma}\hat{M}_\theta^{-1}\|\sum_j\|\hat{a}_j\|^2=O_\mathbb{P}(1)$. Since $\rho_j$ is nonincreasing in $j$, the min-max principle gives $\hat{\tau}_k\leq \bar{C}_n\rho_k^2$ for every $k$. Bounding the integrand in $p_\Delta$ by one on $\{W<1/2\}$ and using $1-\prod_k(1-x_k)\leq\sum_kx_k$ for $x_k\in[0,1]$ and $2(1-\Phi(x))\leq e^{-x^2/2}$ for $x\geq0$, we obtain
    \begin{align*}
    p_\Delta&\leq\mathbb{P}_W(W<1/2)+\sum_{k:\,\hat{\tau}_k>0}2\Big(1-\Phi\big(T_\Delta/(2\sqrt{\hat{\tau}_k})\big)\Big)\\
    &\leq \mathbb{P}_W(W<1/2)+\Upsilon\big(T_\Delta/(2\sqrt{\bar{C}_n})\big),
    \end{align*}
    where $\Upsilon(t):=\sum_{k\geq1}\exp\{-t^2/(2\rho_k^2)\}$. Let $c:=\int_0^\infty\sqrt{\log(M(\epsilon)+1)}\,d\epsilon$, which is finite by Condition \ref{C:decay}. Since $M$ is nonincreasing, $\epsilon\sqrt{\log(M(\epsilon)+1)}\leq c$, so at most $M(2^{-l-1})\leq\exp(c^2 4^{l+1})$ indices $k$ satisfy $2^{-l-1}\leq\rho_k<2^{-l}$, and each contributes at most $\exp(-t^24^l/2)$ to $\Upsilon(t)$. Hence $\Upsilon(t)<\infty$ for $t\geq3c$, and $\Upsilon(t)\rightarrow0$ as $t\rightarrow\infty$ by dominated convergence. Since $\mathbb{P}_W(W<1/2)\xrightarrow{p}0$ and $T_\Delta/\sqrt{\bar{C}_n}\xrightarrow{p}\infty$, $p_\Delta\xrightarrow{p}0$.

    (iii) Let $\zeta_{n,K}$ consist of the first $K_\Delta$ coordinates of $\zeta_n$. Then $p_{\Delta,K}$ is a function of $(\zeta_{n,K}/\hat{\sigma}_n,\hat{\Sigma}_{\Delta,K})$, which converges in distribution to $(\xi,\Sigma_{\Delta,K})$ with $\xi\sim N(0,\Sigma_{\Delta,K})$ by (i) and \Cref{Lem:psi-hat}. Let $\tau_1>\cdots>\tau_r>0$ be the positive eigenvalues of $\Sigma_{\Delta,K}$ with unit eigenvectors $u_1,\ldots,u_r$, and let $F(t):=\prod_{k\leq r}(2\Phi(t/\sqrt{\tau_k})-1)$. Since these eigenvalues are simple, Weyl's inequality and the $\sin\theta$ theorem show that this function is continuous at $(z,\Sigma_{\Delta,K})$ for every nonzero $z$ in the range of $\Sigma_{\Delta,K}$, with value $1-F(\max_{k\leq r}|u_k^Tz|)$: the coordinates of $z$ along eigenvectors with eigenvalues tending to zero vanish in the limit, and the corresponding factors in \cref{eq:cal-T} tend to one. This function is also continuous in $W$ at $W=1$. As $\xi$ lies in this set almost surely, the $u_k^T\xi$ are independent $N(0,\tau_k)$, and $W$ concentrates at one, the continuous mapping theorem shows that $p_{\Delta,K}$ converges in distribution to $1-F(\max_{k\leq r}|u_k^T\xi|)$, which is uniform on $(0,1)$.
\end{proof}

\section{Simulation settings and global tests}\label{sec:G}

This appendix gives the settings of \Cref{subsec:design}, the implementation of the competing tests and checks of the implementations, the rejection rates shown in \Cref{fig:comparison}, and results at other levels and for other computer models.

\subsection{Settings}\label{sec:G-settings}

In the one-dimensional block, $y^s(x)=e^x$, the inputs follow the $N(0.5,0.2^2)$ distribution truncated to $[0,1]$, and $\sigma=0.5$. Before normalization to $\|\delta_0\|_{L_2(p)}=1$, the six discrepancy shapes are the linear trend $x-0.5$, the sine wave $\sin(2\pi x)$, the narrow and wide bumps $\exp\{-(x-0.7)^2/(2b^2)\}$ with $b=0.05$ and $0.15$, the step $\mathbb{I}(x>0.6)$, and the boundary layer $\exp(-x/0.1)$. In the two-dimensional block, the inputs are uniform on $[0,1]^2$, $y^s(x)=1+0.5(x_1+x_2)+0.3x_1x_2+0.2\sin(2\pi x_1)\cos(2\pi x_2)$, and the shapes are the bump $\exp[-\{(x_1-0.7)^2+(x_2-0.7)^2\}/(2\cdot0.1^2)]$, the step $\mathbb{I}(x_1+x_2>1.2)$, and the product $\sin(2\pi x_1)\sin(2\pi x_2)$. For the simple machine, the discrepancy $0.65x/(1+cx)-0.65x$ is not normalized; the loss coefficient $c$ is the root of $\sqrt{n}\,\|\delta\|_{L_2(p)}/\sigma=2.80$. The norms are computed by the midpoint rule with 4001 points in one dimension and $401^2$ points in two dimensions.

\subsection{Competing tests}\label{sec:G-competitors}

In each replication, all tests use the same data, and EH, GLR, and NMCT are applied to the residuals $r_i=y_i-y^s(x_i)$. The versions with an estimated calibration parameter are described in \Cref{sec:I}.

The order-selection test of \citet{eubank1992testing} assumes a null model with $p\geq1$ estimated coefficients. Because $H_0$ has none, we use $p=0$: the cosine basis, including the constant, is orthonormalized at the design points, and $H_0$ is rejected when the order selected by their criterion is positive. The critical values are $c_\alpha=6.74$, $4.18$, and $3.22$ for $\alpha=0.01$, $0.05$, and $0.10$ (their Table~1); by their Theorem~3.1, they do not depend on $p$. The original test with $p=1$, whose basis is orthogonal to the constant, is reported as EH ($p=1$).

The GLR test of \citet{fanjq2001} is applied to the residuals as in \citet[Section 3.4]{fanjq2007}, so that the smoothed function is zero under $H_0$. For the local linear smoother $H_h$ with the Epanechnikov kernel (a product kernel in two dimensions) and bandwidth $h$, let $\lambda(h)=(n/2)\log(\mathrm{RSS}_0/\mathrm{RSS}_1)$, with $\mathrm{RSS}_0=\sum_i r_i^2$ and $\mathrm{RSS}_1=\sum_i\{r_i-(H_hr)_i\}^2$. The bandwidths $h=0.10$, $0.15$, and $0.225$ times the side length of $\Omega$ are combined by the multiscale statistic $T=\max_h\{2\lambda(h)-d(h)\}/\sqrt{2d(h)}$, with $d(h)=\operatorname{tr}(H_h+H_h^T-H_h^TH_h)$. This is Eq.~3.21 of \citet{fanjq2007} with their $r_K$ replaced by 2 and their $\mu_n(h)$ by $d(h)$. The $p$-value comes from their conditional bootstrap (Section~4.1) with 500 draws: the residuals of the alternative fit with the middle bandwidth are centered and resampled, and $p=(1+\#\{T_b\geq T\})/501$.

The NMCT of \citet{zhu2005nonparametric} uses the Kolmogorov--Smirnov statistic $D_n=\sup_x|R_n(x)|$ of the residual-marked empirical process $R_n(x)=n^{-1/2}\sum_i r_i\mathbb{I}(x_i\leq x)$ \citep{stute1997nonparametric}, with the componentwise order in two dimensions. Its null distribution is approximated by random weighting \citep[Section 1.2.3]{zhu2005nonparametric}, $R_n^{(b)}(x)=n^{-1/2}\sum_i e_i^{(b)}r_i\mathbb{I}(x_i\leq x)$, with 500 draws of i.i.d.\ weights that take the values $-(\sqrt{5}-1)/2$ and $(\sqrt{5}+1)/2$ with probabilities $(\sqrt{5}+1)/(2\sqrt{5})$ and $(\sqrt{5}-1)/(2\sqrt{5})$.

The test of \citet{neumeyer2003nonparametric} compares two unknown regression curves, so it needs a second sample: in each replication, $n$ simulator outputs without noise at inputs drawn independently from the design density. We use their statistic $K_N^{(1)}$, local linear estimators with the Epanechnikov kernel, their rule-of-thumb bandwidths with the difference-based variance estimator of \citet{rice1984bandwidth}, and the wild bootstrap with the weights above and 200 draws.

To check the implementation of NMCT, we reproduced Table~1 of \citet{stute1998bootstrap}: the data come from $y=5x+ax^2+e$ with $x\sim U(0,1)$ and $e\sim N(0,\sigma^2)$, the null model $m(x)=\theta x$ is fitted by least squares, and the statistic is $D_n$. For a model linear in $\theta$, NMCT coincides with their wild bootstrap \citep[Section 1.2.3]{zhu2005nonparametric}. In \Cref{tab:G-nmct}, the two sets of rejection rates differ by less than two Monte Carlo standard errors, except under $H_0$ with $n=100$, where our rate of 0.043 is closer to the nominal level than their 0.061.

For GLR, no published table of rejection rates matches our setting, so we checked the Wilks phenomenon \citep{fanjq2001}: under $H_0$, $r_K\lambda(h)$ is approximately $\chi^2$ distributed with $\mu_n+2$ degrees of freedom \citep[Section 3.5]{fanjq2007}. With uniform inputs on $[0,1]$, $n=200$, $N(0,1)$ errors, and 1000 replications, the sample mean and variance of $r_K\lambda(h)$ are 11.56 and 23.98 for $h=0.1$, against $\mu_n+2=11.52$ and $2(\mu_n+2)=23.04$, and 6.29 and 11.86 for $h=0.2$, against 6.76 and 13.52.

\subsection{Rejection rates}\label{sec:G-rates}

\Cref{tab:G-size} gives the Type~I error rates at $\alpha=0.01$, $0.05$, and $0.10$, \Cref{tab:G-power} the rejection rates at $\alpha=0.05$ shown in \Cref{fig:comparison}, and \Cref{tab:G-power01,tab:G-power10} the power at $\alpha=0.01$ and $0.10$. The original EH test with $p=1$ also exceeds the nominal level in most settings.

Under a fully specified $H_0$, EH, GLR, and NMCT depend on the data only through the residuals, which are the same for every $y^s$. FMMT depends on $y^s$ through $C_n$ and $\hat{\sigma}_n$, which are computed from the KRR fit to $y$. \Cref{tab:G-ys} repeats the one-dimensional block with $y^s(x)=1$ and $y^s(x)=\sin(2\pi x)+x$, with the same discrepancies and errors. The Type~I error rates are between 0.047 and 0.055, and the power differs across the three models by less than 0.1, for the linear trend with $n=25$.

\begin{table}[!htbp]
\centering
\caption{Rejection rates of our NMCT implementation and of the wild bootstrap test $D_W$ in Table~1 of \citet{stute1998bootstrap} at $\alpha=0.05$ (1000 replications, 500 bootstrap draws).}
\label{tab:G-nmct}
\begin{tabular}{lccccc}
\toprule
$\sigma^2$ & $a$ & $n$ & Ours & Stute et al. & MC s.e. \\
\midrule
1 & 0 & 25 & 0.051 & 0.065 & 0.008 \\
1 & 0 & 50 & 0.056 & 0.064 & 0.008 \\
1 & 0 & 100 & 0.043 & 0.061 & 0.008 \\
1 & 5 & 25 & 0.599 & 0.586 & 0.016 \\
1 & 5 & 50 & 0.898 & 0.891 & 0.010 \\
1 & 5 & 100 & 0.997 & 0.994 & 0.002 \\
2 & 5 & 25 & 0.348 & 0.352 & 0.015 \\
2 & 5 & 50 & 0.629 & 0.637 & 0.015 \\
2 & 5 & 100 & 0.911 & 0.904 & 0.009 \\
\bottomrule
\end{tabular}
\end{table}

\begin{table}[!htbp]
\centering
\caption{Type~I error rates at three levels (1000 replications). EH: the order-selection test with $p=0$ used in \Cref{sec:simulation}; EH ($p=1$): the original test of \citet{eubank1992testing}.}
\label{tab:G-size}
\begin{tabular}{lccccccc}
\toprule
Setting & $\alpha$ & FMMT & EH & EH ($p=1$) & GLR & NMCT & ND \\
\midrule
1D, $n=25$ & 0.01 & 0.010 & 0.019 & 0.015 & 0.007 & 0.007 & 0.018 \\
1D, $n=25$ & 0.05 & 0.055 & 0.074 & 0.066 & 0.043 & 0.051 & 0.060 \\
1D, $n=25$ & 0.10 & 0.102 & 0.133 & 0.131 & 0.086 & 0.105 & 0.114 \\
1D, $n=50$ & 0.01 & 0.009 & 0.011 & 0.015 & 0.012 & 0.007 & 0.009 \\
1D, $n=50$ & 0.05 & 0.051 & 0.065 & 0.057 & 0.040 & 0.056 & 0.051 \\
1D, $n=50$ & 0.10 & 0.109 & 0.135 & 0.106 & 0.091 & 0.121 & 0.116 \\
Simple machine, $n=25$ & 0.01 & 0.014 & 0.023 & 0.033 & 0.009 & 0.010 & 0.022 \\
Simple machine, $n=25$ & 0.05 & 0.047 & 0.075 & 0.091 & 0.048 & 0.048 & 0.083 \\
Simple machine, $n=25$ & 0.10 & 0.110 & 0.145 & 0.147 & 0.091 & 0.091 & 0.155 \\
Simple machine, $n=50$ & 0.01 & 0.011 & 0.016 & 0.025 & 0.010 & 0.011 & 0.022 \\
Simple machine, $n=50$ & 0.05 & 0.054 & 0.068 & 0.074 & 0.059 & 0.053 & 0.076 \\
Simple machine, $n=50$ & 0.10 & 0.108 & 0.130 & 0.125 & 0.102 & 0.098 & 0.130 \\
\bottomrule
\end{tabular}
\end{table}

\begin{table}[!htbp]
\centering
\caption{Rejection rates at $\alpha=0.05$ shown in \Cref{fig:comparison} (1000 replications; two-dimensional block with $n=200$). EH and ND are defined only in one dimension.}
\label{tab:G-power}
\begin{tabular}{lcccccc}
\toprule
Setting & FMMT & EH & EH ($p=1$) & GLR & NMCT & ND \\
\midrule
Linear trend, $n=25$ & 0.398 & 0.559 & 0.731 & 0.381 & 0.113 & 0.125 \\
Sine wave, $n=25$ & 0.364 & 0.492 & 0.690 & 0.385 & 0.118 & 0.166 \\
Narrow bump, $n=25$ & 0.237 & 0.382 & 0.336 & 0.303 & 0.179 & 0.143 \\
Wide bump, $n=25$ & 0.504 & 0.622 & 0.354 & 0.384 & 0.451 & 0.429 \\
Step, $n=25$ & 0.374 & 0.497 & 0.460 & 0.338 & 0.215 & 0.155 \\
Boundary layer, $n=25$ & 0.304 & 0.336 & 0.387 & 0.308 & 0.144 & 0.086 \\
Linear trend, $n=50$ & 0.437 & 0.541 & 0.762 & 0.480 & 0.134 & 0.114 \\
Sine wave, $n=50$ & 0.428 & 0.518 & 0.722 & 0.465 & 0.166 & 0.141 \\
Narrow bump, $n=50$ & 0.274 & 0.379 & 0.344 & 0.388 & 0.197 & 0.139 \\
Wide bump, $n=50$ & 0.526 & 0.659 & 0.378 & 0.471 & 0.487 & 0.437 \\
Step, $n=50$ & 0.443 & 0.529 & 0.439 & 0.399 & 0.259 & 0.126 \\
Boundary layer, $n=50$ & 0.350 & 0.365 & 0.391 & 0.396 & 0.171 & 0.088 \\
Simple machine, $n=25$ & 0.502 & 0.621 & 0.409 & 0.335 & 0.415 & 0.447 \\
Simple machine, $n=50$ & 0.569 & 0.655 & 0.440 & 0.428 & 0.449 & 0.478 \\
2D, $H_0$ & 0.055 & — & — & 0.046 & 0.064 & — \\
2D, Bump & 0.202 & — & — & 0.132 & 0.174 & — \\
2D, Step & 0.326 & — & — & 0.128 & 0.197 & — \\
2D, Sine $\times$ sine & 0.215 & — & — & 0.118 & 0.050 & — \\
\bottomrule
\end{tabular}
\end{table}

\begin{table}[!htbp]
\centering
\caption{Power at $\alpha=0.01$ (1000 replications).}
\label{tab:G-power01}
\begin{tabular}{lcccccc}
\toprule
Setting & FMMT & EH & EH ($p=1$) & GLR & NMCT & ND \\
\midrule
Linear trend, $n=25$ & 0.161 & 0.307 & 0.541 & 0.175 & 0.012 & 0.027 \\
Sine wave, $n=25$ & 0.134 & 0.250 & 0.499 & 0.166 & 0.015 & 0.044 \\
Narrow bump, $n=25$ & 0.085 & 0.191 & 0.164 & 0.136 & 0.054 & 0.052 \\
Wide bump, $n=25$ & 0.260 & 0.420 & 0.192 & 0.178 & 0.187 & 0.203 \\
Step, $n=25$ & 0.161 & 0.272 & 0.255 & 0.144 & 0.088 & 0.065 \\
Boundary layer, $n=25$ & 0.137 & 0.167 & 0.210 & 0.166 & 0.021 & 0.029 \\
Linear trend, $n=50$ & 0.175 & 0.274 & 0.578 & 0.230 & 0.034 & 0.029 \\
Sine wave, $n=50$ & 0.169 & 0.236 & 0.507 & 0.233 & 0.038 & 0.047 \\
Narrow bump, $n=50$ & 0.089 & 0.177 & 0.170 & 0.176 & 0.070 & 0.041 \\
Wide bump, $n=50$ & 0.270 & 0.394 & 0.194 & 0.236 & 0.244 & 0.208 \\
Step, $n=50$ & 0.176 & 0.251 & 0.260 & 0.201 & 0.080 & 0.044 \\
Boundary layer, $n=50$ & 0.168 & 0.154 & 0.207 & 0.248 & 0.056 & 0.027 \\
Simple machine, $n=25$ & 0.284 & 0.410 & 0.218 & 0.138 & 0.178 & 0.226 \\
Simple machine, $n=50$ & 0.310 & 0.398 & 0.232 & 0.213 & 0.219 & 0.253 \\
\bottomrule
\end{tabular}
\end{table}

\begin{table}[!htbp]
\centering
\caption{Power at $\alpha=0.10$ (1000 replications).}
\label{tab:G-power10}
\begin{tabular}{lcccccc}
\toprule
Setting & FMMT & EH & EH ($p=1$) & GLR & NMCT & ND \\
\midrule
Linear trend, $n=25$ & 0.557 & 0.683 & 0.815 & 0.529 & 0.223 & 0.211 \\
Sine wave, $n=25$ & 0.497 & 0.634 & 0.771 & 0.509 & 0.239 & 0.260 \\
Narrow bump, $n=25$ & 0.354 & 0.503 & 0.457 & 0.441 & 0.300 & 0.240 \\
Wide bump, $n=25$ & 0.627 & 0.728 & 0.455 & 0.521 & 0.594 & 0.544 \\
Step, $n=25$ & 0.504 & 0.610 & 0.560 & 0.453 & 0.325 & 0.248 \\
Boundary layer, $n=25$ & 0.405 & 0.449 & 0.486 & 0.380 & 0.246 & 0.147 \\
Linear trend, $n=50$ & 0.610 & 0.703 & 0.833 & 0.586 & 0.261 & 0.190 \\
Sine wave, $n=50$ & 0.571 & 0.644 & 0.803 & 0.597 & 0.308 & 0.250 \\
Narrow bump, $n=50$ & 0.386 & 0.502 & 0.471 & 0.510 & 0.293 & 0.230 \\
Wide bump, $n=50$ & 0.660 & 0.762 & 0.480 & 0.600 & 0.615 & 0.537 \\
Step, $n=50$ & 0.568 & 0.631 & 0.566 & 0.530 & 0.364 & 0.211 \\
Boundary layer, $n=50$ & 0.476 & 0.488 & 0.500 & 0.500 & 0.283 & 0.138 \\
Simple machine, $n=25$ & 0.621 & 0.726 & 0.508 & 0.461 & 0.543 & 0.544 \\
Simple machine, $n=50$ & 0.689 & 0.751 & 0.539 & 0.561 & 0.578 & 0.577 \\
\bottomrule
\end{tabular}
\end{table}

\begin{table}[!htbp]
\centering
\caption{Rejection rates of FMMT at $\alpha=0.05$ in the one-dimensional block with three computer models $y^s$ and the same discrepancies and errors (1000 replications).}
\label{tab:G-ys}
\begin{tabular}{lccc}
\toprule
Setting & $y^s(x)=e^x$ & $y^s(x)=1$ & $y^s(x)=\sin(2\pi x)+x$ \\
\midrule
$H_0$, $n=25$ & 0.055 & 0.052 & 0.047 \\
Linear trend, $n=25$ & 0.398 & 0.417 & 0.325 \\
Sine wave, $n=25$ & 0.364 & 0.367 & 0.327 \\
Narrow bump, $n=25$ & 0.237 & 0.239 & 0.226 \\
Wide bump, $n=25$ & 0.504 & 0.516 & 0.458 \\
Step, $n=25$ & 0.374 & 0.377 & 0.347 \\
Boundary layer, $n=25$ & 0.304 & 0.302 & 0.305 \\
$H_0$, $n=50$ & 0.051 & 0.049 & 0.048 \\
Linear trend, $n=50$ & 0.437 & 0.449 & 0.404 \\
Sine wave, $n=50$ & 0.428 & 0.430 & 0.421 \\
Narrow bump, $n=50$ & 0.274 & 0.279 & 0.261 \\
Wide bump, $n=50$ & 0.526 & 0.541 & 0.477 \\
Step, $n=50$ & 0.443 & 0.438 & 0.415 \\
Boundary layer, $n=50$ & 0.350 & 0.343 & 0.353 \\
\bottomrule
\end{tabular}
\end{table}

\section{Subdomain tests}\label{sec:H}

This appendix extends \Cref{subsec:subdomain} to other partitions, to the one-dimensional case, and to other corrections for multiple testing.

In the one-dimensional block, the inputs are uniform on $[0,1]$, $y^s(x)=e^x$, $\sigma=0.5$, and $n=50$ or $100$. The partitions consist of $N=2$, $3$, $4$, or $6$ subdomains of equal length. For $N=3$, we also shift both cut points $1/3$ and $2/3$ by $-0.10$, $-0.05$, $0.05$, or $0.10$. The discrepancies are the bump $\cos^2\{\pi(x-0.15)/0.2\}$ on $[0.05,0.25]$ and $\cos(12\pi x)$ on $[0,1/3)$, both zero elsewhere, normalized as in \Cref{subsec:design} and taken at the strength $c$ of \Cref{subsec:design} or $2c$. The two-dimensional block is that of \Cref{subsec:subdomain}, with the $2\times2$ and $3\times3$ grids. A subdomain carries the discrepancy if the discrepancy does not vanish on it, and we report the probability that at least one such subdomain is rejected. Each setting is replicated 1000 times.

Under $H_0$ (\Cref{tab:H-fwer}), the family-wise error rate with the Bonferroni correction is between 0.038 and 0.060 for the equal partitions and between 0.047 and 0.058 for the shifted cut points. Without a correction, it grows with the number of subdomains, from 0.087 to 0.242. The procedures of \citet{vsidak1967rectangular} and \citet{hochberg1988sharper} give nearly the same rates as the Bonferroni correction. Under the alternatives below, replacing the Bonferroni correction by the procedure of \citet{holm1979simple}, \v{S}id\'ak, or Hochberg raises the rates in \Cref{tab:H-power1d,tab:H-power2d,tab:H-shift} by at most 0.01. Under these alternatives, a subdomain without the discrepancy is also rejected more often than under $H_0$, because the KRR estimate uses the data on both sides of a subdomain boundary. With the Bonferroni correction, the probability that at least one such subdomain is rejected is at most 0.04 for $\cos(12\pi x)$ and 0.09 for the local bump at strength $c$; at $2c$, it is at most 0.10 for the local bump, except with $N=4$, whose first cut point 0.25 is the end of the support of the bump, where it is 0.16 with $n=50$ and 0.17 with $n=100$. In two dimensions, it is 0.045 at $c$ and 0.090 at $2c$ with the $2\times2$ grid and at most 0.041 with the $3\times3$ grid.

For the local bump (\Cref{tab:H-power1d}), the subdomain tests detect the discrepancy more often than the global test for every partition. For $\cos(12\pi x)$ on $[0,1/3)$, neither the global test nor the subdomain tests detect the discrepancy at strength $c$. At $2c$, the subdomain tests detect it much more often than the global test, for example 0.593 against 0.177 with $N=3$ and $n=100$. In two dimensions (\Cref{tab:H-power2d}), the $2\times2$ grid, one of whose subdomains coincides with the support of the discrepancy, gives higher rates than the $3\times3$ grid. With shifted cut points (\Cref{tab:H-shift}), the rates change considerably, for example from 0.289 to 0.593 for $\cos(12\pi x)$ at $2c$ with $n=100$. The subdomains should therefore be chosen from physical knowledge before the data are examined (\Cref{sec:glo_sub_tests}).

\begin{table}[!htbp]
\centering
\caption{Family-wise error rates of the subdomain tests under $H_0$ at $\alpha=0.05$ (1000 replications). Shift: $N=3$ with both cut points shifted by the amount shown. Under $H_0$, the procedure of Holm gives the same family-wise error rate as the Bonferroni correction.}
\label{tab:H-fwer}
\begin{tabular}{lcccc}
\toprule
Setting & None & Bonferroni & \v{S}id\'ak & Hochberg \\
\midrule
1D, $n=50$, $N=2$ & 0.087 & 0.040 & 0.041 & 0.042 \\
1D, $n=50$, $N=3$ & 0.127 & 0.053 & 0.054 & 0.053 \\
1D, $n=50$, $N=4$ & 0.152 & 0.038 & 0.038 & 0.039 \\
1D, $n=50$, $N=6$ & 0.181 & 0.040 & 0.041 & 0.040 \\
1D, $n=50$, shift $-0.10$ & 0.114 & 0.049 & 0.049 & 0.050 \\
1D, $n=50$, shift $-0.05$ & 0.130 & 0.053 & 0.055 & 0.053 \\
1D, $n=50$, shift $+0.05$ & 0.120 & 0.051 & 0.051 & 0.051 \\
1D, $n=50$, shift $+0.10$ & 0.121 & 0.047 & 0.047 & 0.048 \\
1D, $n=100$, $N=2$ & 0.104 & 0.048 & 0.048 & 0.050 \\
1D, $n=100$, $N=3$ & 0.148 & 0.058 & 0.059 & 0.060 \\
1D, $n=100$, $N=4$ & 0.183 & 0.054 & 0.055 & 0.055 \\
1D, $n=100$, $N=6$ & 0.219 & 0.060 & 0.062 & 0.060 \\
1D, $n=100$, shift $-0.10$ & 0.143 & 0.058 & 0.059 & 0.058 \\
1D, $n=100$, shift $-0.05$ & 0.142 & 0.058 & 0.060 & 0.059 \\
1D, $n=100$, shift $+0.05$ & 0.151 & 0.058 & 0.062 & 0.060 \\
1D, $n=100$, shift $+0.10$ & 0.150 & 0.051 & 0.052 & 0.053 \\
2D, $n=200$, $2\times2$ & 0.150 & 0.044 & 0.045 & 0.044 \\
2D, $n=200$, $3\times3$ & 0.242 & 0.046 & 0.046 & 0.046 \\
\bottomrule
\end{tabular}
\end{table}

\begin{table}[!htbp]
\centering
\caption{Rejection rates of the global test and probability that a subdomain carrying the discrepancy is rejected with the Bonferroni correction, one-dimensional block (1000 replications). The local bump is supported on $[0.05,0.25]$ and the high-frequency discrepancy is $\cos(12\pi x)$ on $[0,1/3)$.}
\label{tab:H-power1d}
\begin{tabular}{lccccc}
\toprule
Setting & Global & $N=2$ & $N=3$ & $N=4$ & $N=6$ \\
\midrule
Local bump, $c$, $n=50$ & 0.238 & 0.275 & 0.313 & 0.292 & 0.253 \\
Local bump, $c$, $n=100$ & 0.227 & 0.279 & 0.323 & 0.291 & 0.268 \\
Local bump, $2c$, $n=50$ & 0.603 & 0.745 & 0.803 & 0.799 & 0.741 \\
Local bump, $2c$, $n=100$ & 0.685 & 0.826 & 0.887 & 0.878 & 0.841 \\
High frequency, $c$, $n=50$ & 0.071 & 0.070 & 0.063 & 0.069 & 0.061 \\
High frequency, $c$, $n=100$ & 0.063 & 0.053 & 0.061 & 0.059 & 0.067 \\
High frequency, $2c$, $n=50$ & 0.104 & 0.275 & 0.320 & 0.250 & 0.343 \\
High frequency, $2c$, $n=100$ & 0.177 & 0.505 & 0.593 & 0.431 & 0.593 \\
\bottomrule
\end{tabular}
\end{table}

\begin{table}[!htbp]
\centering
\caption{As \Cref{tab:H-power1d} for the two-dimensional block with $n=200$ and the discrepancy $\sin(2\pi x_1)\sin(2\pi x_2)$ on $[0.5,1]^2$.}
\label{tab:H-power2d}
\begin{tabular}{lccc}
\toprule
Setting & Global & $2\times2$ & $3\times3$ \\
\midrule
Quadrant, $c$ & 0.226 & 0.349 & 0.265 \\
Quadrant, $2c$ & 0.767 & 0.953 & 0.889 \\
\bottomrule
\end{tabular}
\end{table}

\begin{table}[!htbp]
\centering
\caption{As \Cref{tab:H-power1d} for $N=3$ with both cut points $1/3$ and $2/3$ shifted by the amount shown.}
\label{tab:H-shift}
\begin{tabular}{lccccc}
\toprule
Setting & $-0.10$ & $-0.05$ & $0$ & $+0.05$ & $+0.10$ \\
\midrule
Local bump, $c$, $n=50$ & 0.338 & 0.313 & 0.313 & 0.278 & 0.253 \\
Local bump, $c$, $n=100$ & 0.341 & 0.333 & 0.323 & 0.284 & 0.253 \\
Local bump, $2c$, $n=50$ & 0.821 & 0.832 & 0.803 & 0.735 & 0.713 \\
Local bump, $2c$, $n=100$ & 0.897 & 0.904 & 0.887 & 0.844 & 0.814 \\
High frequency, $c$, $n=50$ & 0.081 & 0.070 & 0.063 & 0.054 & 0.054 \\
High frequency, $c$, $n=100$ & 0.070 & 0.071 & 0.061 & 0.046 & 0.037 \\
High frequency, $2c$, $n=50$ & 0.257 & 0.286 & 0.320 & 0.203 & 0.178 \\
High frequency, $2c$, $n=100$ & 0.411 & 0.453 & 0.593 & 0.302 & 0.289 \\
\bottomrule
\end{tabular}
\end{table}

\section{Tests with calibration parameters}\label{sec:I}

This appendix gives the details of \Cref{subsec:calibration} and studies the interaction between calibration and validation.

\subsection{Settings, competing tests, and estimates of \texorpdfstring{$\theta$}{theta}}\label{sec:I-settings}

In the simple machine of \Cref{subsec:calibration}, calibration absorbs the component of the discrepancy $\delta_c(x)=0.65x/(1+cx)-0.65x$ along the gradient $x$, so the loss coefficient $c$ is chosen so that the remainder $\delta_c-\Pi\delta_c$, where $\Pi$ is the $L_2(p)$ projection onto the span of $x$, satisfies $\sqrt{n}\,\|\delta_c-\Pi\delta_c\|_{L_2(p)}/\sigma=2.80$. This gives $c=0.042$ for $n=50$, close to the value $1/20$ of \citet{brynjarsdottir2014learning}.

In the exponential example of \Cref{subsec:calibration}, the inputs on $[0,1]$ have density $p(x)=0.5+x$, $\sigma=0.1$, and $\Theta=[0.1,10]\times[-5,5]$. For the models linear in $\theta$, here and in \Cref{sec:I-interaction}, $\hat{\theta}$ is computed without bounds, because the minimizer in \cref{eq:thetastar} is unique and finite. The discrepancy is $s\,d(x)$, where $d(x)=b(x)(a_0+a_1x+a_2x^2)$, $b(x)=\sin^6\{3\pi(x-2/3)\}$ on $[2/3,1]$ and zero elsewhere, and $(a_0,a_1,a_2)$ is chosen so that $d$ is orthogonal in $L_2$ to $\partial y^s/\partial\theta_1$ and $\partial y^s/\partial\theta_2$ at $\theta$ and $\|d\|_{L_2}=1$. Hence $\theta^*=\theta$, and $H_0$ holds on $[0,2/3]$. \Cref{subsec:calibration} uses $s=0.3$.

The naive FMMT applies the test of \Cref{sec:methodology} to the computer model $y^s(\cdot,\hat{\theta})$ as if it were known. GLR and NMCT estimate $\theta$ by least squares over the same $\Theta$. For GLR, the residuals $r_i=y_i-y^s(x_i,\hat{\theta})$ enter the statistic of \Cref{sec:G-competitors}, and the bootstrap samples $y^s(x_i,\hat{\theta})+\varepsilon_i^*$ are refitted, $\theta$ included, in every draw \citep[Section 4.1]{fanjq2007}. For NMCT, the random weights act on the linearized summands \citep[Remark 1.2.1]{zhu2005nonparametric},
\[
R_n^{(b)}(x)=n^{-1/2}\sum_j e_j^{(b)}r_j\{\mathbb{I}(x_j\leq x)-L(x_j)^TG(x)\},
\]
where $L(x)=A^{-1}\nabla_\theta y^s(x,\hat{\theta})$, $A=n^{-1}\sum_i\nabla_\theta y^s(x_i,\hat{\theta})\nabla_\theta y^s(x_i,\hat{\theta})^T$, and $G(x)=n^{-1}\sum_i\nabla_\theta y^s(x_i,\hat{\theta})\mathbb{I}(x_i\leq x)$. For a model linear in $\theta$, this is the wild bootstrap of \citet{stute1998bootstrap}, whose results we reproduced in \Cref{sec:G-competitors}.

\Cref{tab:I-theta} summarizes $\hat{\theta}$. Under $H_0$, the estimates are centered at the true values. Under the frictional loss, the mean of $\hat{\theta}$ is 0.577 for $n=50$ and 0.600 for $n=100$, because calibration absorbs most of the loss. In the exponential example with the bump and $n=50$, $\hat{\theta}$ varies much more than in the other settings, and in 6 of the 1000 replications it lies on the boundary of $\Theta$. Its median is $(1.001,0.797)$.

\subsection{Interaction between calibration and validation}\label{sec:I-interaction}

Because $\hat{\theta}$ is estimated from the data on the whole domain, a discrepancy on one subdomain also affects the tests on the others. We place the discrepancy $s\,d(x)$ of \Cref{sec:I-settings} on the third of three subdomains of equal length, with $s=0$, $0.05$, $0.1$, and $0.3$ and $n=100$ and $200$. Besides the exponential family, we use the linear family $y^s(x,\theta)=\theta_1+\theta_2x$ with $\theta=(1,0.5)$, the same design, and the same $\sigma$; for this family, $(a_0,a_1,a_2)$ is chosen so that $d$ is orthogonal to $1$ and $x$. \Cref{tab:I-interaction} reports the rejection rates with the Bonferroni correction.

Under $H_0$, the rejection rates of the global test are between 0.020 and 0.046, and the subdomain tests are conservative, with rates of at most 0.006 on subdomain 3 and 0.015 on subdomains 1 and 2. When the discrepancy is weak, the test on the third subdomain detects it more often than the global test, for example 0.526 against 0.268 for the linear family with $s=0.05$ and $n=100$. On the first two subdomains, where the model with $\theta^*$ is correct, the family-wise error rate is larger when $s>0$, at most 0.140 for $n=100$ and 0.069 for $n=200$. One reason is that the discrepancy increases the variability of $\hat{\theta}$: for the exponential family with $n=100$, the standard deviation of $\hat{\theta}_1$ rises from 0.021 at $s=0$ to 0.098 at $s=0.3$. The additional variability comes from the error of KRR in approximating the discrepancy, which changes with the random inputs. This error vanishes asymptotically and is therefore not part of $\Sigma_\Delta$ in \Cref{Th:cal}, but it vanishes slowly when the discrepancy is sharp.

\begin{table}[!htbp]
\centering
\caption{Mean (standard deviation) of $\hat{\theta}$ over 1000 replications. Simple machine: $y^s(x,\theta)=\theta x$, true value 0.65; exponential: $y^s(x,\theta)=\theta_1\exp(\theta_2x)$, true value $(1,0.8)$.}
\label{tab:I-theta}
\begin{tabular}{lccc}
\toprule
Setting & $n$ & $\theta$ or $\theta_1$ & $\theta_2$ \\
\midrule
Simple machine, $H_0$ & 50 & 0.650 (0.008) & — \\
Simple machine, $H_0$ & 100 & 0.650 (0.005) & — \\
Simple machine, frictional loss & 50 & 0.577 (0.007) & — \\
Simple machine, frictional loss & 100 & 0.600 (0.004) & — \\
Exponential, $H_0$ & 50 & 1.000 (0.031) & 0.800 (0.043) \\
Exponential, $H_0$ & 100 & 1.001 (0.020) & 0.799 (0.027) \\
Exponential, local bump & 50 & 1.055 (0.578) & 0.757 (0.454) \\
Exponential, local bump & 100 & 1.000 (0.050) & 0.803 (0.068) \\
\bottomrule
\end{tabular}
\end{table}

\begin{table}[!htbp]
\centering
\caption{Calibrated tests when the discrepancy, of $L_2$ norm $s$, lies on the third of three subdomains (1000 replications, $\alpha=0.05$, Bonferroni correction). Subdomain 3: rejection rate on the subdomain carrying the discrepancy; FWER: probability of rejecting on subdomain 1 or 2. s.d.\ of $\hat{\theta}$: standard deviations of $\hat{\theta}_1$ and $\hat{\theta}_2$.}
\label{tab:I-interaction}
\begin{tabular}{lccccc}
\toprule
Family & $s$ & Global & Subdomain 3 & FWER & s.d.\ of $\hat{\theta}$ \\
\midrule
Linear, $n=100$ & 0 & 0.023 & 0.006 & 0.013 & 0.028, 0.044 \\
Linear, $n=100$ & 0.05 & 0.268 & 0.526 & 0.063 & 0.051, 0.077 \\
Linear, $n=100$ & 0.1 & 0.888 & 0.946 & 0.109 & 0.070, 0.105 \\
Linear, $n=100$ & 0.3 & 0.967 & 0.996 & 0.116 & 0.069, 0.106 \\
Linear, $n=200$ & 0 & 0.020 & 0.004 & 0.007 & 0.017, 0.027 \\
Linear, $n=200$ & 0.05 & 0.730 & 0.952 & 0.047 & 0.021, 0.033 \\
Linear, $n=200$ & 0.1 & 0.993 & 0.999 & 0.069 & 0.028, 0.043 \\
Linear, $n=200$ & 0.3 & 0.998 & 1.000 & 0.061 & 0.028, 0.043 \\
Exponential, $n=100$ & 0 & 0.037 & 0.006 & 0.008 & 0.021, 0.029 \\
Exponential, $n=100$ & 0.05 & 0.271 & 0.540 & 0.059 & 0.032, 0.043 \\
Exponential, $n=100$ & 0.1 & 0.904 & 0.959 & 0.109 & 0.062, 0.067 \\
Exponential, $n=100$ & 0.3 & 0.990 & 0.998 & 0.140 & 0.098, 0.103 \\
Exponential, $n=200$ & 0 & 0.046 & 0.004 & 0.015 & 0.014, 0.019 \\
Exponential, $n=200$ & 0.05 & 0.727 & 0.974 & 0.041 & 0.017, 0.023 \\
Exponential, $n=200$ & 0.1 & 0.993 & 1.000 & 0.062 & 0.017, 0.023 \\
Exponential, $n=200$ & 0.3 & 0.999 & 1.000 & 0.067 & 0.019, 0.026 \\
\bottomrule
\end{tabular}
\end{table}

\section{Sensitivity to tuning choices, the basis, and the design}\label{sec:J}

This appendix reports the sensitivity of FMMT to its tuning choices, the orthonormal basis, the design, and the adjustment for the estimated $\sigma$, summarized in \Cref{subsec:other}. In \Cref{sec:J-tuning}, we use the one-dimensional block of \Cref{subsec:design} with $n=50$ and 1000 replications, and add the high-frequency discrepancy $\cos(12\pi x)$, normalized in the same way, at the strengths $c$ and $2c$. Its frequency 6 lies above $k_{\max}=3$ and below the default $k_{\max}=\lfloor\sqrt{50}\rfloor=7$. One choice at a time departs from the defaults $\nu=3.5$, $\phi=1$, $\ell=0.7$, and $k_{\max}=7$ of \Cref{sec:details}; the other choices are fixed, and the same data sets are used throughout.

\subsection{Tuning parameters and the basis}\label{sec:J-tuning}

\Cref{tab:J-tuning} shows that the Type~I error rate stays between 0.042 and 0.058. Apart from two exceptions, the power changes by at most 0.03. First, a larger $\ell$, which makes $\rho_j$ decay faster, lowers the power for all six shapes, most for the linear trend and the sine wave: with $\ell=1.5$, it falls by 0.25 and 0.24 for these two and by about 0.05 for the step and the boundary layer. A larger $\ell$ raises $\rho_1/\rho_j$ for every $j\ge2$, for example $\rho_1/\rho_2=(\log3/\log2)^\ell$ from 1.38 at $\ell=0.7$ to 2.00 at $\ell=1.5$, which puts more weight on $\hat{s}_n^{(1)}$, whereas $s^{(1)}=0$ for these two shapes, because they are odd about 0.5 and the design density is symmetric about 0.5. Second, a smaller $\nu$ raises the power against $\cos(12\pi x)$ at $2c$, from 0.07 with $\nu=3.5$ to 0.34 with $\nu=1.5$, because a less smooth kernel shrinks high frequencies less. The truncation $k_{\max}$ has little effect, because the KRR estimate already shrinks the frequencies near $k_{\max}$.

\Cref{tab:J-cv} compares the default $\nu=3.5$ with $\nu$ chosen together with $C_n$ by five-fold cross-validation over $\nu\in\{1.5,2.5,3.5,5.5\}$. The Type~I error rate is 0.054, the power changes by at most 0.01 for the six shapes of \Cref{subsec:design}, and the power against $\cos(12\pi x)$ at $2c$ rises from 0.07 to 0.31, as cross-validation selects $\nu=1.5$ in about half of these replications.

\Cref{tab:J-basis} replaces the Fourier basis by the cosine basis $\{1,\sqrt{2}\cos(\pi kx)\}$, the orthonormalized Legendre polynomials, the periodized Daubechies-4 wavelets, and the eigenfunctions of the integral operator with kernel $\sqrt{p(x)}K(x,u)\sqrt{p(u)}$. Each basis is ordered as listed (wavelets by level, eigenfunctions by decreasing eigenvalue), its $j$th function receives the decay coefficient of the $j$th Fourier function, and the same number of functions is used, rounded up to a power of 2 for the wavelets. The Type~I error rate is between 0.047 and 0.049 for every basis. No basis is the most powerful for every shape. The wavelets give the highest power for the narrow bump, the wide bump, and the step and the lowest for the linear trend. The kernel eigenfunctions give the highest power for $\cos(12\pi x)$ at $2c$ and the lowest for the step and the boundary layer.

\subsection{Designs and the estimated \texorpdfstring{$\sigma$}{sigma}}\label{sec:J-designs}

\Cref{tab:J-random,tab:J-fixed} report rejection rates with $n=50$ in one dimension and $n=196$, the $14\times14$ grid, in two dimensions, with $y^s$ as in \Cref{subsec:design} and the discrepancies normalized under the limiting density of each design. In \Cref{tab:J-random}, the Latin hypercube and the scrambled Sobol$'$ designs are redrawn in each replication; in \Cref{tab:J-fixed}, every design is fixed. The one-dimensional grid and quantile design satisfy the conditions of \Cref{rem:fixed}. Across these designs, the Type~I error rate of FMMT is between 0.035 and 0.065, and the power is within 0.07 of that under the i.i.d.\ designs.

\Cref{tab:J-sigma} reports the Type~I error rates of the global test with and without the adjustment of \Cref{sec:details}, in the null settings of \Cref{subsec:design} with the data of the main comparison. Without the adjustment, the rates at $\alpha=0.05$ are 0.069 to 0.089 instead of 0.047 to 0.055 in one dimension, and 0.060 instead of 0.055 in two dimensions.

\begin{table}[!htbp]
\centering
\caption{Rejection rates of FMMT at $\alpha=0.05$ when one tuning choice departs from the default $\nu=3.5$, $\phi=1$, $\ell=0.7$, $k_{\max}=7$ (one-dimensional block of \Cref{subsec:design}, $n=50$, 1000 replications). HF: $\cos(12\pi x)$ at the strength $c$ of \Cref{subsec:design}; HF ($2c$): at $2c$.}
\label{tab:J-tuning}
\small\setlength{\tabcolsep}{4pt}
\begin{tabular}{lccccccccc}
\toprule
Setting & $H_0$ & Lin. & Sine & N.\ bump & W.\ bump & Step & B.\ layer & HF & HF ($2c$) \\
\midrule
Default & 0.049 & 0.443 & 0.421 & 0.242 & 0.529 & 0.412 & 0.352 & 0.060 & 0.073 \\
$\nu=1.5$ & 0.042 & 0.421 & 0.430 & 0.254 & 0.533 & 0.423 & 0.334 & 0.080 & 0.343 \\
$\nu=2.5$ & 0.048 & 0.429 & 0.423 & 0.247 & 0.535 & 0.403 & 0.347 & 0.064 & 0.136 \\
$\nu=5.5$ & 0.057 & 0.434 & 0.419 & 0.246 & 0.540 & 0.412 & 0.357 & 0.065 & 0.066 \\
$\phi=0.5$ & 0.045 & 0.416 & 0.424 & 0.254 & 0.522 & 0.409 & 0.338 & 0.069 & 0.093 \\
$\phi=2$ & 0.049 & 0.447 & 0.415 & 0.237 & 0.546 & 0.413 & 0.356 & 0.064 & 0.086 \\
$\ell=0.6$ & 0.051 & 0.471 & 0.448 & 0.237 & 0.535 & 0.409 & 0.357 & 0.054 & 0.081 \\
$\ell=1.0$ & 0.058 & 0.346 & 0.323 & 0.232 & 0.527 & 0.394 & 0.329 & 0.064 & 0.069 \\
$\ell=1.5$ & 0.055 & 0.190 & 0.179 & 0.228 & 0.514 & 0.363 & 0.307 & 0.062 & 0.070 \\
$k_{\max}=3$ & 0.051 & 0.445 & 0.423 & 0.243 & 0.531 & 0.412 & 0.350 & 0.060 & 0.063 \\
$k_{\max}=14$ & 0.049 & 0.443 & 0.421 & 0.242 & 0.529 & 0.412 & 0.353 & 0.059 & 0.073 \\
\bottomrule
\end{tabular}
\end{table}

\begin{table}[!htbp]
\centering
\caption{Rejection rates of FMMT with $\nu=3.5$ and with $\nu$ chosen by cross-validation (CV) together with $C_n$, and the proportion of replications in which each $\nu$ is chosen (same data as \Cref{tab:J-tuning}).}
\label{tab:J-cv}
\small
\begin{tabular}{lcccccc}
\toprule
 & \multicolumn{2}{c}{Rejection rate} & \multicolumn{4}{c}{Proportion of $\nu$ chosen by CV} \\
\cmidrule(lr){2-3}\cmidrule(lr){4-7}
Shape & $\nu=3.5$ & CV & $\nu=1.5$ & $\nu=2.5$ & $\nu=3.5$ & $\nu=5.5$ \\
\midrule
$H_0$ & 0.049 & 0.054 & 0.12 & 0.08 & 0.10 & 0.70 \\
Lin. & 0.443 & 0.436 & 0.10 & 0.08 & 0.09 & 0.73 \\
Sine & 0.421 & 0.424 & 0.13 & 0.08 & 0.09 & 0.70 \\
N.\ bump & 0.242 & 0.251 & 0.37 & 0.10 & 0.10 & 0.43 \\
W.\ bump & 0.529 & 0.538 & 0.19 & 0.12 & 0.08 & 0.61 \\
Step & 0.412 & 0.419 & 0.25 & 0.11 & 0.10 & 0.54 \\
B.\ layer & 0.352 & 0.348 & 0.12 & 0.10 & 0.10 & 0.68 \\
HF & 0.060 & 0.080 & 0.24 & 0.07 & 0.09 & 0.60 \\
HF ($2c$) & 0.073 & 0.310 & 0.52 & 0.11 & 0.08 & 0.29 \\
\bottomrule
\end{tabular}
\end{table}

\begin{table}[!htbp]
\centering
\caption{As \Cref{tab:J-tuning}, with other orthonormal bases and the default tuning choices.}
\label{tab:J-basis}
\small\setlength{\tabcolsep}{4pt}
\begin{tabular}{lccccccccc}
\toprule
Basis & $H_0$ & Lin. & Sine & N.\ bump & W.\ bump & Step & B.\ layer & HF & HF ($2c$) \\
\midrule
Fourier & 0.049 & 0.443 & 0.421 & 0.242 & 0.529 & 0.412 & 0.352 & 0.060 & 0.073 \\
Cosine & 0.048 & 0.468 & 0.370 & 0.234 & 0.521 & 0.398 & 0.361 & 0.063 & 0.073 \\
Legendre & 0.049 & 0.465 & 0.377 & 0.242 & 0.527 & 0.402 & 0.371 & 0.064 & 0.066 \\
Wavelet & 0.047 & 0.417 & 0.399 & 0.333 & 0.590 & 0.443 & 0.338 & 0.061 & 0.110 \\
Eigenfunctions & 0.047 & 0.451 & 0.382 & 0.258 & 0.558 & 0.373 & 0.314 & 0.056 & 0.140 \\
\bottomrule
\end{tabular}
\end{table}

\begin{table}[!htbp]
\centering
\caption{Rejection rates of FMMT at $\alpha=0.05$ when the Latin hypercube and the scrambled Sobol$'$ designs are redrawn in each replication (1000 replications; $n=50$ in one dimension and $n=196$ in two). The i.i.d.\ designs are for comparison. Bump: the narrow bump in one dimension.}
\label{tab:J-random}
\small
\begin{tabular}{lccc}
\toprule
Design & $H_0$ & Bump & Step \\
\midrule
1D, Latin hypercube & 0.064 & 0.247 & 0.474 \\
1D, Sobol$'$ & 0.051 & 0.274 & 0.502 \\
1D, i.i.d.\ uniform & 0.050 & 0.277 & 0.475 \\
1D, i.i.d.\ truncated normal & 0.053 & 0.267 & 0.445 \\
2D, Latin hypercube & 0.050 & 0.147 & 0.347 \\
2D, Sobol$'$ & 0.049 & 0.193 & 0.347 \\
2D, i.i.d.\ uniform & 0.061 & 0.168 & 0.332 \\
\bottomrule
\end{tabular}
\end{table}

\begin{table}[!htbp]
\centering
\caption{As \Cref{tab:J-random}, with every design fixed across replications: the midpoint grid, the quantile design $x_i=F^{-1}\{(i-1/2)/n\}$ of the truncated normal distribution, one draw of a Latin hypercube, and the unscrambled Sobol$'$ and Halton sequences.}
\label{tab:J-fixed}
\small
\begin{tabular}{lccc}
\toprule
Design & $H_0$ & Bump & Step \\
\midrule
1D, grid & 0.051 & 0.234 & 0.491 \\
1D, quantile & 0.056 & 0.264 & 0.443 \\
1D, Latin hypercube & 0.046 & 0.234 & 0.467 \\
1D, Sobol$'$ & 0.044 & 0.213 & 0.494 \\
1D, Halton & 0.065 & 0.222 & 0.420 \\
2D, grid & 0.055 & 0.170 & 0.380 \\
2D, Latin hypercube & 0.035 & 0.176 & 0.370 \\
2D, Sobol$'$ & 0.042 & 0.182 & 0.352 \\
2D, Halton & 0.035 & 0.168 & 0.311 \\
\bottomrule
\end{tabular}
\end{table}

\begin{table}[!htbp]
\centering
\caption{Type~I error rates of the global FMMT with / without the adjustment for the estimated $\sigma$ (null settings of \Cref{subsec:design}, the data of the main comparison, 1000 replications).}
\label{tab:J-sigma}
\small
\begin{tabular}{lcccc}
\toprule
Setting & $n$ & $\alpha=0.01$ & $\alpha=0.05$ & $\alpha=0.10$ \\
\midrule
One-dimensional & 25 & 0.010 / 0.021 & 0.055 / 0.088 & 0.102 / 0.139 \\
One-dimensional & 50 & 0.009 / 0.014 & 0.051 / 0.069 & 0.109 / 0.125 \\
Simple machine & 25 & 0.014 / 0.023 & 0.047 / 0.089 & 0.110 / 0.160 \\
Simple machine & 50 & 0.011 / 0.017 & 0.054 / 0.081 & 0.108 / 0.133 \\
Two-dimensional & 200 & 0.016 / 0.021 & 0.055 / 0.060 & 0.101 / 0.107 \\
\bottomrule
\end{tabular}
\end{table}

\section{The shear-layer application}\label{sec:K}

This appendix gives the checks summarized in \Cref{sec:case_study}.

\subsection{A known simulator}\label{sec:K-simulator}

To have a simulator that is known exactly, we fit the four-parameter logistic curve $g(M)=d+(a-d)/\{1+(M/c)^b\}$ to the 11 runs by least squares, which gives $a=1.006$, $d=0.322$, $c=0.898$, and $b=2.98$. The curve $g$ differs from the runs by at most 0.031. The kernel interpolant $\hat{g}$ of the values of $g$ at the 11 run inputs, computed as in \Cref{sec:case_study}, differs from $g$ by at most 0.0003 on $[0.1,1.5]$, by at most 0.0046 on $[0,0.1)$, where it extrapolates, and by at most 0.0011 at the 32 physical inputs. By \Cref{rem:surrogate}, with $\rho_1=1.29$ and $\hat{\sigma}_n=0.088$, replacing $g$ by $\hat{g}$ changes the test statistic by at most $1.29\times\sqrt{32}\times0.0011/0.088\approx0.09$, whereas the observed global statistic is 4.32.

\subsection{Type I error rates}\label{sec:K-size}

We generated 1000 data sets $y_i=g(x_i)+e_i$ at the 32 observed inputs, with $e_i\sim N(0,0.088^2)$, so that $H_0$ holds. Each data set is tested in three ways: against $\hat{g}$ with the asymptotic $p$-values of \Cref{sec:methodology}; against $\hat{g}$ with a parametric bootstrap with 200 draws, in which $y^*=\hat{g}(x)+\hat{\sigma}_ne^*$ and each draw is refitted, including the cross-validation; and against $g$ itself with the asymptotic $p$-values. In \Cref{tab:K-size}, the asymptotic $p$-values give Type~I error rates of 0.048 for the global test and between 0.034 and 0.044 for the subdomain tests, including the subdomains with two observations. The bootstrap gives rates between 0.040 and 0.050, and the rates with $g$ differ from those with $\hat{g}$ by at most 0.001.

\subsection{Robustness of the conclusions}\label{sec:K-robust}

\Cref{tab:K-real} gives the $p$-values on the physical data with the interpolant of all 11 runs, with the 11 interpolants that each leave out one run, and with a parametric bootstrap with 1000 draws. In all 13 cases, the global test rejects, and, with the Bonferroni correction, the subdomain tests reject on subdomains 2 to 5 and not on subdomains 1 and 6. Without the correction, the conclusion at level 0.05 changes only for subdomain 1, where the interpolant extrapolates: its $p$-value falls from 0.18 to 0.047 when the run at $M_c=0.10$ is left out.

\begin{table}[!htbp]
\centering
\caption{Type~I error rates at $\alpha=0.05$ when the data are generated from the logistic simulator $g$ at the 32 observed inputs (1000 replications; unadjusted for the subdomains). Interpolant: $\hat{g}$, asymptotic or parametric bootstrap $p$-values (200 draws); true: $g$, asymptotic $p$-values.}
\label{tab:K-size}
\small
\begin{tabular}{lcccc}
\toprule
Test & Observations & Interpolant & Interpolant, bootstrap & True \\
\midrule
Global & 32 & 0.048 & 0.045 & 0.047 \\
Subdomain 1 & 2 & 0.044 & 0.043 & 0.043 \\
Subdomain 2 & 5 & 0.037 & 0.040 & 0.037 \\
Subdomain 3 & 12 & 0.035 & 0.043 & 0.036 \\
Subdomain 4 & 9 & 0.034 & 0.043 & 0.034 \\
Subdomain 5 & 2 & 0.042 & 0.050 & 0.042 \\
Subdomain 6 & 2 & 0.039 & 0.050 & 0.039 \\
\bottomrule
\end{tabular}
\end{table}

\begin{table}[!htbp]
\centering
\caption{Unadjusted $p$-values on the shear-layer data with the interpolant of all runs, with the interpolants that leave out the run at the value of $M_c$ shown, and with a parametric bootstrap (1000 draws, so the smallest attainable value is $1/1001$). With the Bonferroni correction, a subdomain is rejected when its $p$-value is at most $0.05/6\approx0.0083$.}
\label{tab:K-real}
\small\setlength{\tabcolsep}{3pt}
\begin{tabular}{lccccccc}
\toprule
Interpolant & Global & S1 & S2 & S3 & S4 & S5 & S6 \\
\midrule
All runs & $2.8\times10^{-3}$ & 0.176 & $3.2\times10^{-3}$ & $1.1\times10^{-4}$ & $2.8\times10^{-7}$ & $4.3\times10^{-4}$ & 0.251 \\
Without 0.10 & $4.0\times10^{-3}$ & 0.047 & $1.6\times10^{-3}$ & $9.1\times10^{-5}$ & $2.6\times10^{-7}$ & $4.2\times10^{-4}$ & 0.232 \\
Without 0.24 & $2.4\times10^{-3}$ & 0.111 & $2.1\times10^{-3}$ & $8.5\times10^{-5}$ & $2.6\times10^{-7}$ & $4.5\times10^{-4}$ & 0.240 \\
Without 0.38 & $3.7\times10^{-3}$ & 0.178 & $4.0\times10^{-3}$ & $1.5\times10^{-4}$ & $3.0\times10^{-7}$ & $4.3\times10^{-4}$ & 0.264 \\
Without 0.52 & $1.5\times10^{-3}$ & 0.086 & $1.7\times10^{-3}$ & $5.7\times10^{-5}$ & $3.1\times10^{-7}$ & $3.6\times10^{-4}$ & 0.233 \\
Without 0.66 & $3.9\times10^{-3}$ & 0.148 & $1.9\times10^{-3}$ & $8.2\times10^{-5}$ & $1.3\times10^{-7}$ & $5.6\times10^{-4}$ & 0.200 \\
Without 0.80 & $3.6\times10^{-3}$ & 0.205 & $5.5\times10^{-3}$ & $2.6\times10^{-4}$ & $1.1\times10^{-6}$ & $5.7\times10^{-4}$ & 0.418 \\
Without 0.94 & $1.5\times10^{-3}$ & 0.148 & $2.3\times10^{-3}$ & $5.1\times10^{-5}$ & $5.6\times10^{-8}$ & $1.2\times10^{-4}$ & 0.199 \\
Without 1.08 & $4.0\times10^{-3}$ & 0.169 & $3.0\times10^{-3}$ & $1.2\times10^{-4}$ & $4.2\times10^{-7}$ & $9.1\times10^{-4}$ & 0.222 \\
Without 1.22 & $2.7\times10^{-3}$ & 0.177 & $3.3\times10^{-3}$ & $1.1\times10^{-4}$ & $2.7\times10^{-7}$ & $3.8\times10^{-4}$ & 0.295 \\
Without 1.36 & $3.4\times10^{-3}$ & 0.173 & $3.1\times10^{-3}$ & $1.0\times10^{-4}$ & $2.7\times10^{-7}$ & $5.8\times10^{-4}$ & 0.232 \\
Without 1.50 & $2.6\times10^{-3}$ & 0.175 & $3.2\times10^{-3}$ & $1.1\times10^{-4}$ & $3.0\times10^{-7}$ & $4.1\times10^{-4}$ & 0.232 \\
Bootstrap & $2.0\times10^{-3}$ & 0.126 & $4.0\times10^{-3}$ & $1.0\times10^{-3}$ & $1.0\times10^{-3}$ & $1.0\times10^{-3}$ & 0.175 \\
\bottomrule
\end{tabular}
\end{table}

\putbib
\end{bibunit}

\end{document}